\documentclass[useAMS,usenatbib]{mn2e}

\usepackage{graphicx}

\def\ltorder{\mathrel{\raise.3ex\hbox{$<$}\mkern-14mu
             \lower0.6ex\hbox{$\sim$}}}
\def\aap{{A\&A}}		
\def\aapr{{A\&A~Rev.}} 
\def\apj{{ApJ}}
\def\apjl{{ApJ}}		
\def\aj{{AJ}}
\def\apss{{Ap\&SS}} 
\def\mnras{{MNRAS}}
\def\msun{M_\odot}

\title[Monte Carlo Simulations of Star Clusters - IV. Calibration and Comparison
 with Observations]{Monte Carlo Simulations of Star Clusters - IV. Calibration 
 of the  Monte Carlo Code and Comparison with Observations for  the Open Cluster M67}
\author[M. Giersz, D.C. Heggie \& J.R. Hurley]{Mirek Giersz$^{1}$\thanks{E-mail:
mig@camk.edu.pl (MG); d.c.heggie@ed.ac.uk (DCH)}, Douglas C.
Heggie$^{2}$ and Jarrod R. Hurley$^{3}$
\\
$^{1}$Nicolaus Copernicus Astronomical Centre, Polish Academy of Sciences, 
ul. Bartycka 18, 00-716 Warsaw, Poland\\
$^{2}$University of Edinburgh, School of Mathematics and Maxwell
  Institute for Mathematical Sciences, King's
Buildings,\\ Edinburgh EH9 3JZ, UK\\
$^{3}$Centre for Astrophysics \& Supercomputing, Swinburne University of Technology, Hawthorn VIC 3122, Australia}

\begin{document}

\date{Accepted \ldots. Received \ldots; in original form \ldots}

\pagerange{\pageref{firstpage}--\pageref{lastpage}} \pubyear{2002}

\maketitle

\label{firstpage}

\begin{abstract}
We outline the steps needed in order to incorporate the evolution of
single and binary stars into a particular Monte Carlo code for the
dynamical evolution of a star cluster. We calibrate the results
against $N$-body simulations, and present models for the evolution of
the old open cluster M67 (which has been studied thoroughly in the
literature with $N$-body techniques). The calibration is done by choosing
appropriate free code parameters. We describe in particular the
evolution of the binary, white dwarf and blue straggler populations,
though not all channels for blue straggler formation are represented
yet in our simulations. Calibrated Monte Carlo runs show  good agreement
with results of $N$-body simulations not only for global cluster parameters,
but also for e.g. binary fraction, luminosity function and surface brightness.
Comparison of Monte Carlo simulations with observational data for M67 shows
that is possible to get reasonably good agreement between them.
Unfortunately, because of the large statistical fluctuations of the numerical 
data and uncertainties in the observational data the inferred conclusions 
about the cluster initial conditions are not firm.
\end{abstract}


\section{Introduction}

The modelling of individual globular clusters has a long history (see
\cite{MH1997}, especially \S\S 3 and 7.7).  Much of the
focus of this work is on static models such as the King model and its
variants.  In this kind of modelling the dynamical history of the cluster is
almost irrelevant, except for the general assumption that the cluster
is almost relaxed.  By contrast, there have been a small number of
studies based on techniques which can follow the dynamical evolution
of a cluster.  Most of this work has been performed with a
Fokker-Planck scheme using finite differences,
but also there are examples of the use of fluid and Monte Carlo
methods  (Tab. \ref{table:history}).

\begin{table*}
\begin{minipage}{120mm}
\caption{Dynamical evolutionary models of individual globular clusters}
\begin{tabular}{lll}
\hline
    Cluster&Method&Reference\\
\hline
    47 Tuc&Fokker-Planck&\citet{behleretal2003}, \citet{murphyetal1998}\\
    M15&Fokker-Planck&\citet{murphyetal2003,murphyetal1997,murphyetal1994},
   \citet{dulletal1997}, \citet{grabhornetal1992}\\
    M30&Fokker-Planck&\citet{howelletal2000}\\
    N6397&Fokker-Planck&\citet{dulletal1994}, \citet{drukier1993, drukier1992}\\
    N6624&Fokker-Planck&\citet{grabhornetal1992}\\
    $\omega$ Cen&Monte Carlo&\citet{gh2003}\\
    M3&Moment equations&\citet{angelettietal1980}
\end{tabular}\label{table:history}
    \end{minipage}
  \end{table*}

In the present paper we develop the Monte Carlo technique further, and
apply it to a new object.  The dynamical ingredients of the Monte
Carlo code are essentially the same as those described in
\citet{giersz2006}, whose code embodies several features introduced by
\citet{stod1986}, whose code was in turn based on that originally
devised by \citet{henon1971}.  Three main features distinguish the code
which is described in the present paper from that used by \citet{gh2003} 
in their work on $\omega$ Cen: (i) it now incorporates
dynamical interactions between binary and single stars, between pairs
of binaries, and interactions of three single stars resulting in the
creation of new binaries, all using cross sections; (ii) it replaces
the skeletal approach to stellar evolution taken from \citet{cw1990}
by the algorithms of \citet{hurleyetal2000} for the evolution of single
stars, supplemented by the methods of \citet{hurleyetal2002} for the internal
evolution of binary stars; (iii) a better treatment of the escape process 
in the presence of a static tidal field according to the theory 
proposed by \citet{baum2001}.

There are several factors which motivate this work. Star clusters
are the focus of several intensive observational campaigns 
\citep[e.g.][]{bedinetal2001,bedinetal2003,grindlayetal2001,piottoetal2002,
kaliarietal2003,kafkaetal2004,richeretal2004,andersonetal2006}, which
are now turning to an examination of the parameters of their populations of
binaries and blue stragglers (BS).  Dynamical models are needed 
for the design and interpretation of observational programmes: how is the period
distribution and the spatial distribution of binaries affected by
dynamical evolution?  Another problem is the abundance and spatial
distribution of blue stragglers, which can only be answered by a
technique which follows simultaneously both their dynamics and
internal evolution.   While $N$-body techniques may 
ultimately be the
method of choice for such studies, systems with the size of a globular
cluster are likely to remain beyond reach for some years, simply
because of the number of stars and the population of binaries.  After
all, it is only recently that the ``hardest" open clusters have been
modelled at the necessary level of sophistication, and even then the
typical simulation takes one month \citep{hurleyetal2005}.  These
authors focused on the old open cluster M67, which has been chosen by the MODEST 
international collaboration (MOdelling DEnse STellar systems) \citep{sillsetal2003}
as a target cluster for comparison between observations and
various techniques of numerical simulation. We also focus on
this cluster, partly for the purpose of refining our calibration of
the Monte Carlo method.

This paper begins in Sec.2 with a summary of the features which have
been added to the Monte Carlo scheme.  We also show there how we
calibrate the Monte Carlo technique with $N$-body simulations.  Next
(Sec. 3) we apply the technique to construct a dynamical evolutionary
model of the old open cluster M67, and compare our results with 
observations. We give predictions for the 
initial parameters of the old open cluster M67. The final section 
summarises our conclusions, and discusses some of the main limitations 
of our models.

\section{Technique}

\subsection{Coding of binary- and single-star evolution}

From the dynamical point of view our Monte Carlo code is almost
exactly as described in \citet{giersz2006}. In this technique a star
cluster is treated as a collection of spherical shells, each one
representing a single star with a certain energy and angular momentum.
Neighbouring shells are allowed to interact and exchange energy and
angular momentum at a rate determined by the theory of relaxation.
Escapers are removed according to a prescription which mimics the
effect of a tide.  Shells corresponding to binary stars also
interact with single stars, and other binary stars, at rates
determined by cross sections drawn from the literature \citep{giersz2001}.
The only dynamical alterations deal with tightly bound subsystems,
which often arise in systems with a large mass range (e.g. those
including both stellar-mass black holes and stars at the
hydrogen-burning limit of the main sequence).  

The introduction of stellar and binary evolution has been greatly
facilitated by the development of the ``McScatter" interface
\citep{heggieetal2006}, which provides subroutines for initialising
the stellar evolution of single and binary stars, and for retrieving
the results of subsequent evolution, mass loss, merging of binary
components, etc.  At present two such packages for stellar evolution
can be employed.  One of these is SeBa \citep{pzv1996}, which is
incorporated within the STARLAB environment \citep{hut2003}.  The
other is referred to as ``BSE" (binary star evolution), and is based on
the extensive formulae for the evolution of single stars of a range of
metallicities given by \citet{hurleyetal2000}, along with the
treatment of binaries presented by \citet{hurleyetal2002}.  Most of
our effort has been conducted with BSE, partly to minimise any
development problems with mixed-language programming, and partly
because SeBa is at present restricted to solar metallicity, whereas
our interest is mainly directed to globular clusters.  Generally
speaking, the use of the McScatter interface poses few problems:
\begin{enumerate}
\item There was one instance of a named common block in BSE
which by coincidence was the same as the name of one common block in
the Monte Carlo code; a change of name in the Monte Carlo code was
sufficient cure. 
\item The enumeration of stars requires care.
Although it is not clear from the interface, the numerical identity of
each binary determines the numerical identity of the two single stars
from which it is composed, and it is important that the numerical
identities of all single stars (both binary components and those which
are genuinely single) are different.
\end{enumerate}

\begin{figure}
\includegraphics[angle=-90, width=8cm]{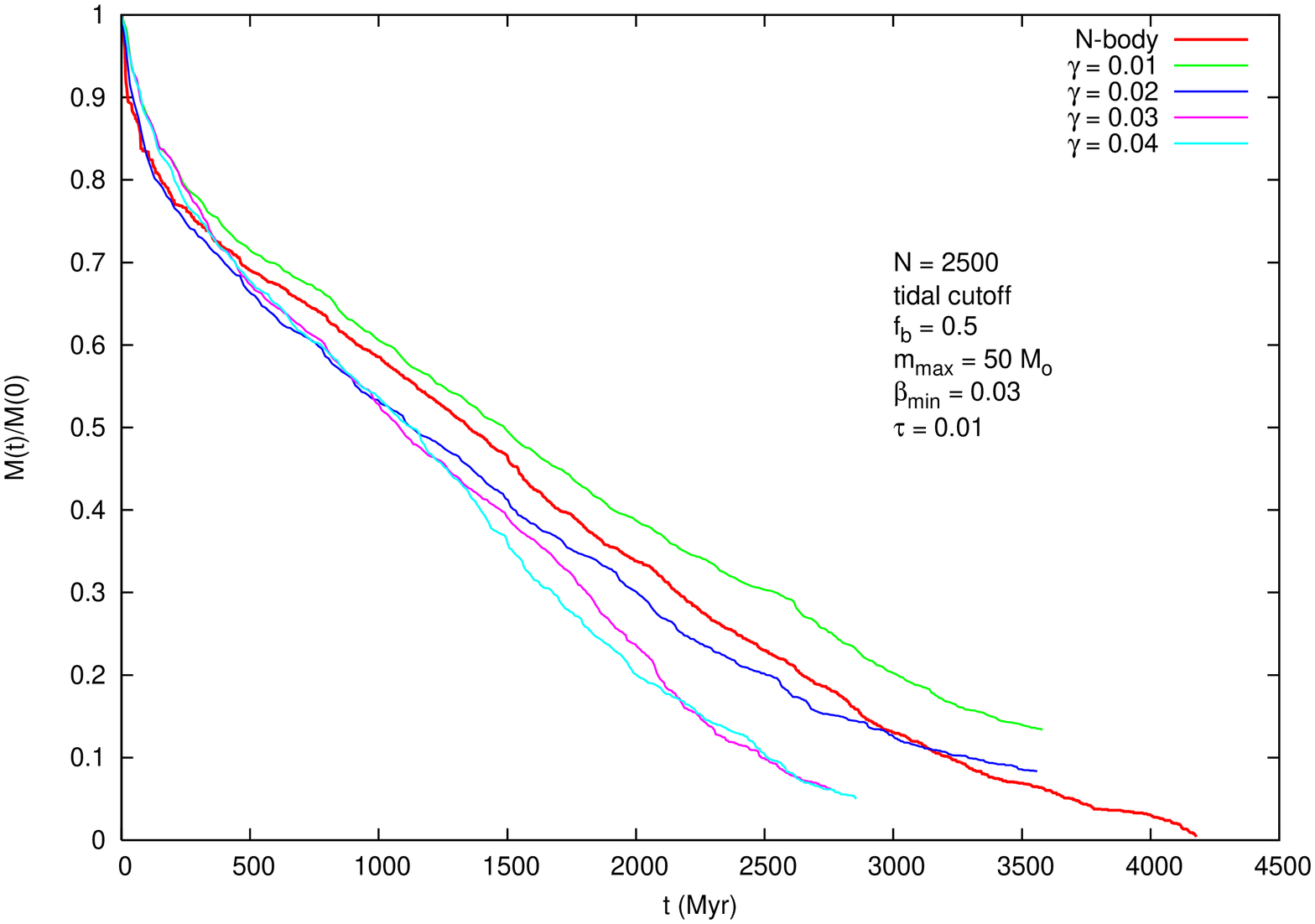}
\includegraphics[angle=-90, width=8cm]{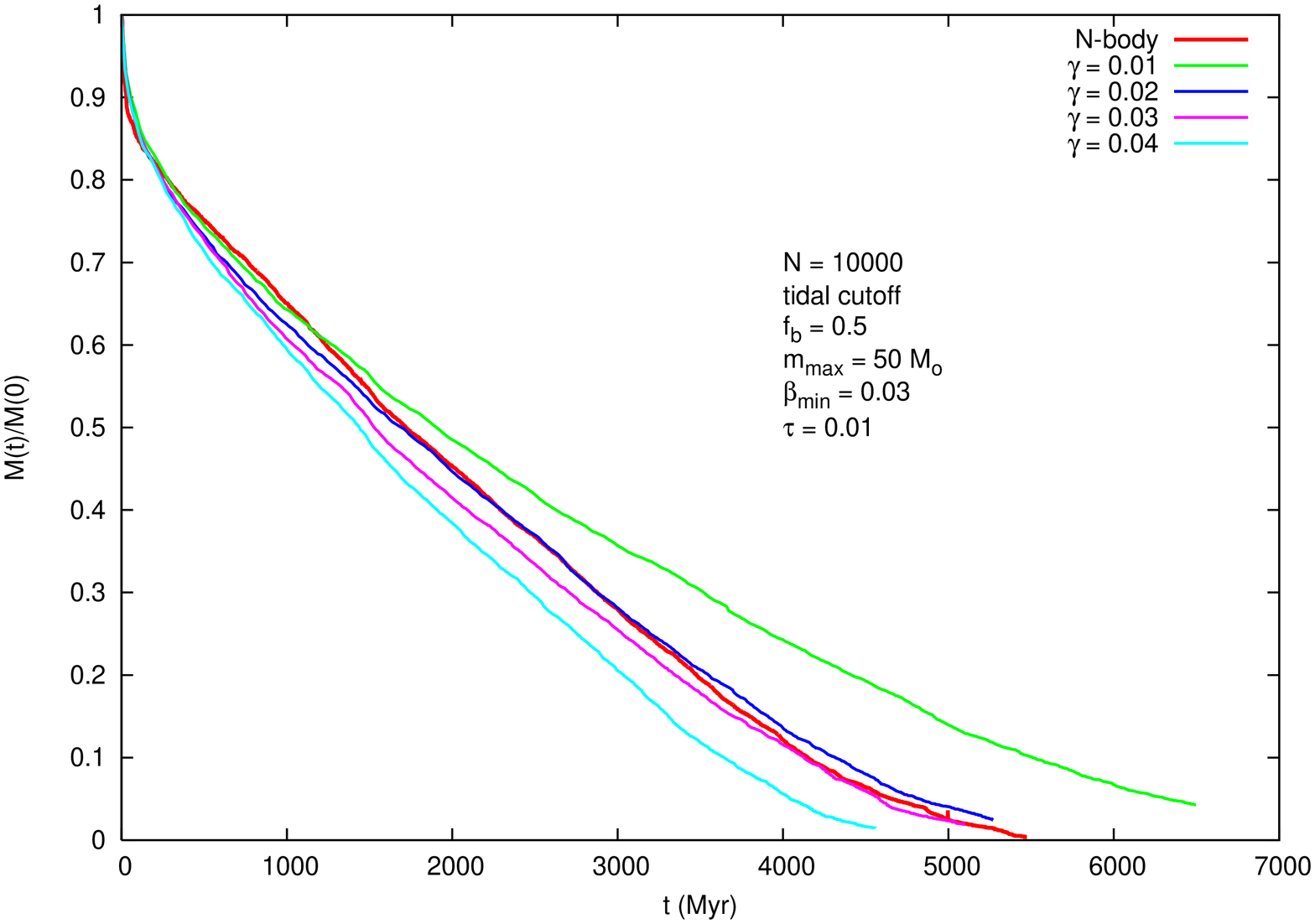}
\caption{Comparison of $N$-body and Monte Carlo simulations for models 
with tidal cutoff. The mass
(normalised by its initial value) is given as a function of time. The
initial conditions are given in Tab. \ref{table:comparison} and described
in the figures. The value of $N$ referred to is $N_s+N_b$, but this 
differs from the total number of particles ($N_s+2N_b$).  The $N$-body 
model is the heavy continuous line, and the others are Monte-Carlo 
simulations with (from the left) $\gamma=0.04,0.03,0.02,0.01$.}
\label{fig:mass_tc_comparison}
\end{figure}

During a time step of the Monte Carlo code, the changes caused by
relaxation and dynamical interactions between binaries and single
stars are performed, and then the stellar evolution of all stars and
binaries is updated.  The associated loss of mass (if any) is
incorporated into the data for each star and binary in the Monte Carlo
code, and any mergers are dealt with by altering the numbers of single
and binary stars and adjusting the parameters of the bodies affected.

\subsection{Calibration}\label{sec:mc}

In a Monte Carlo simulation it is usual to adopt units such that the
constant of gravitation, the initial total mass and the initial virial
radius are 1.  In order to incorporate stellar evolution into a Monte
Carlo simulation, dimensional values for the initial total mass and
virial radius (or equivalent) must be specified, and then the unit of
time in the Monte Carlo code (which is essentially a relaxation
  time) is expressible dimensionally with a factor proportional to
$N/\log(\gamma N)$, where $\gamma$ is a constant and $N$ is the number
of stars in the system.  While the value of $\gamma$ is rather well
known for the case of single stars of equal mass, i.e. $\gamma = 0.11$
approximately \citep{gh1994,joshietal2000}, the case of unequal masses
with a population of primordial binaries has been studied much less.
On analytical grounds \citet{henon1975} gave a formula which, by way
of example, yields a value $\gamma = 0.007$ approximately for a mass
function for single stars of the form $dN\propto m^{-2}dm$, over a
range in which the mass ratio between the maximum and minimum mass is
$500:1$.  This theory also implies that the value depends on the mass
ratio, which changes through stellar evolution.  \citet{gh1996} 
found a value $\gamma\simeq0.015$ for a
power-law mass function of index $-2.5$ and a smaller mass range of
$37.5:1$, by means of intercomparison of $N$-body simulations, and
somewhat larger values from H\'enon's formula or from comparison with
isotropic Fokker-Planck models.

Here we adopt a pragmatic approach, comparing Monte Carlo and $N$-body
models with identical initial conditions.  These are summarised in
Tab. \ref{table:comparison}, where the tidal radius refers to a tidal
{\sl cutoff} (or, for the $N$-body models discussed from
  Sec.\ref{sec:2.2.2} onwards, a tidal field).  It is
necessary to carry out this comparison for at least two values of $N$.
If we were to determine $\gamma$ from a single value of $N$, it might
be that this value simply obscures some systematic problem with the
Monte Carlo code, and would fail for a different value of $N$.  Modest
values of $N$ are better for this purpose, as the $N$-dependence of
the Coulomb logarithm becomes weaker as $N$ increases.  We study cases
with $N = N_s + 2N_b = 3750$ and $15000$, where $N_s, N_b$ are the
initial numbers of single and binary stars, respectively. 

The Monte Carlo code free parameters are as follows: (i) \  $\gamma$, 
(ii) \  $\beta_{min}$, minimum value of the deflection angle 
\citep{giersz1998}, (iii) \  $\tau$, the overall time step and (iv) \ 
$\alpha$, see for the definition Sec. \ref{sec:2.2.2}.

\begin{figure}
\includegraphics[angle=-90, width=8cm]{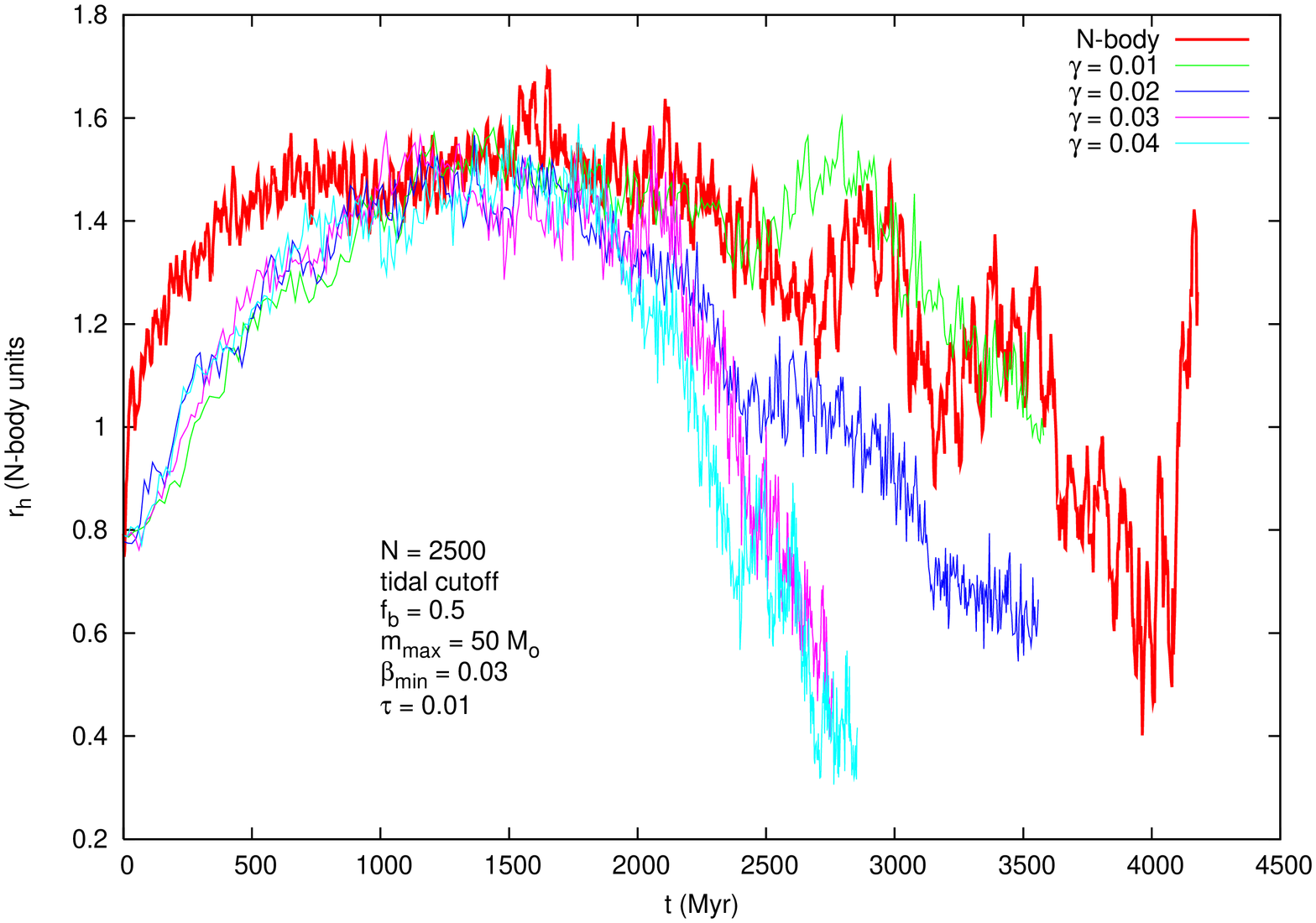}
\includegraphics[angle=-90, width=8cm]{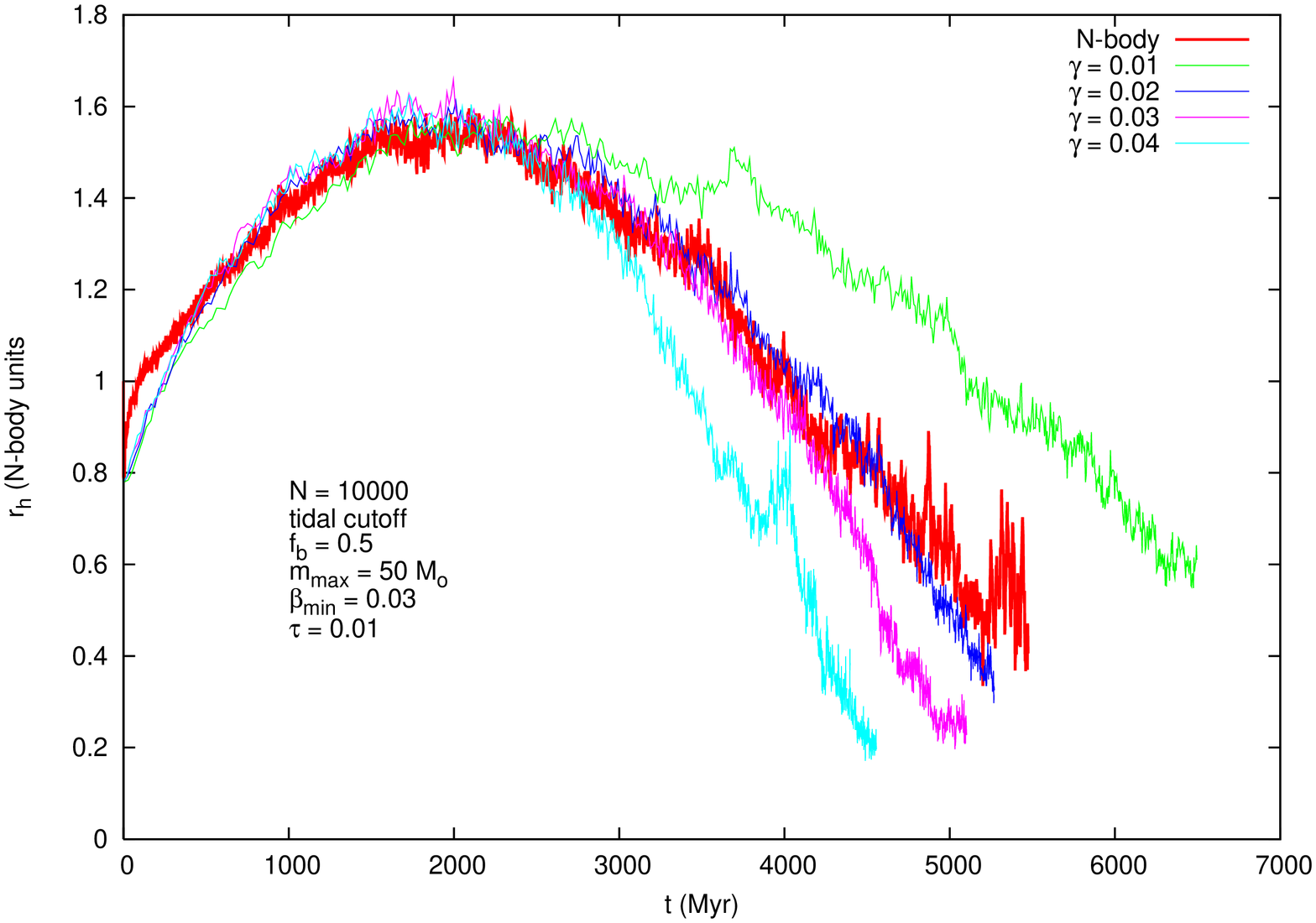}
\caption{Comparison of the evolution of the half-mass radius 
in $N$-body and Monte Carlo simulations for models with tidal cutoff. 
The initial conditions are given in Tab. \ref{table:comparison} 
and described in the figures. The value of $N$ referred to is $N_s+N_b$. 
The $N$-body model is the heavy continuous line, 
and the others are Monte-Carlo simulations with (from the left, at
late times) $\gamma=0.04,0.03,0.02,0.01$. The mismatch at early times in the
upper panel is discussed in \ref{sec:2.2.2}.}
\label{fig:rh_tc_comparison}
\end{figure}

\begin{table}
\caption{Initial conditions of calibration runs}
\begin{tabular}{ll}
	$N_s+2N_b$&$3750$ ($15000$)\\
	Initial model&Plummer\\
	Initial tidal radius&$30$pc ($30$pc)\\
   Initial half-mass radius&$3$pc ($3$pc)\\
	Initial mass function&\citet{kroupaetal1993}\\
	&with $\alpha_1 =
	1.3$, mass range\\
   & between $0.1M_{\odot}$ and $50M_{\odot}$\\
	Binary fraction&$N_s/(N_s+N_b) = 0.5$\\
	Binary eccentricities&$f(e) = 2e$\\
	Binary semi-major axes&Uniformly distributed in the\\
	&
	logarithm in the range\\
	&$2(R_1+R_2)$ to 50AU\\
	Run time (Monte Carlo)&0.4 min (3 min)\\
	Run time (NBODY4 with& 41 min (1400 min)\\
	GRAPE6Af)&
\end{tabular}\label{table:comparison}

Notes: where two values are given, the first value refers to runs with
$N_s+2N_b=3750$, the second (in brackets) to those with $N_s+2N_b = 15000$.
The timings are on a 3GHz PC ($N$-body) and AMD Opetron 242 (Monte Carlo).
\end{table}

\subsubsection{Models with tidal cutoff}\label{sec:2.2.1}

First, we concentrated on calibration of Monte Carlo models for which
the influence of the tidal field of a parent galaxy is characterised 
by the tidal energy cutoff - all stars which have energy larger than 
$E_{t_c} = -GM/r_{t}$ are immediately removed from the system -- $M$ 
is the total mass and $r_{t}$ is the tidal radius. The comparison of 
the evolution of the mass with time (Fig. \ref{fig:mass_tc_comparison}) 
suggests that a value just about $\gamma = 0.02$ is an appropriate
choice (especially for an age of order a few Gyr, as in M67).

Apart from mass, the other fundamental measure of a cluster is its
radius, and the same comparison for the half-mass radius ($r_h$) is presented
in Fig. \ref{fig:rh_tc_comparison}.  It is however, much less
discriminating of the appropriate value of $\gamma$, particularly for 
$N=2500$. A comparison of the two panels also suggests caution in applying the Monte 
Carlo method to a single system with $N\ltorder10^3$, 
because of the increasing role of statistical fluctuations.

To properly assess the inferred values of the free parameters of the 
Monte Carlo code it is important to check the intrinsic statistical 
fluctuation of the code. 
\begin{figure}
\includegraphics[angle=-90, width=8cm]{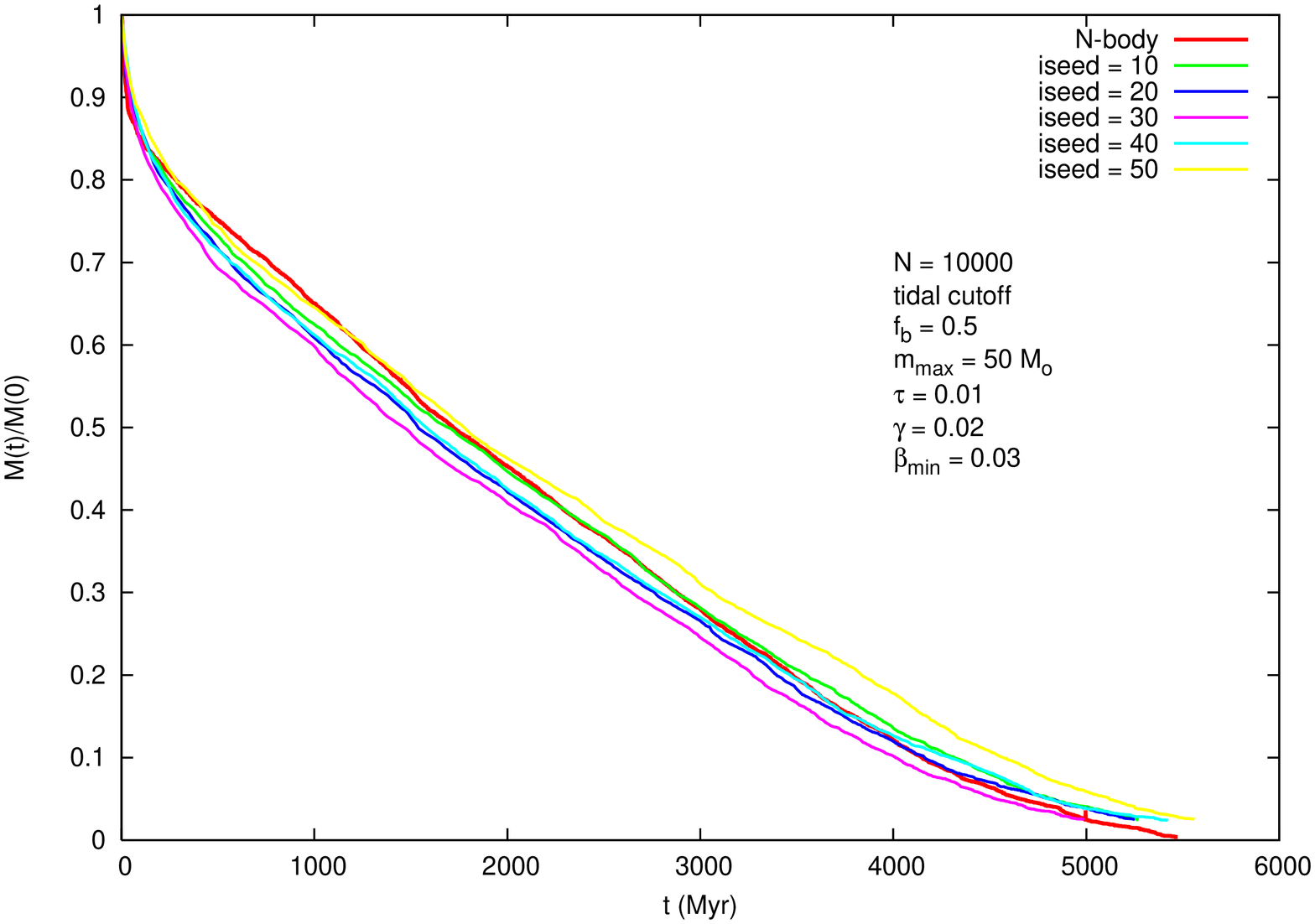}
\includegraphics[angle=-90, width=8cm]{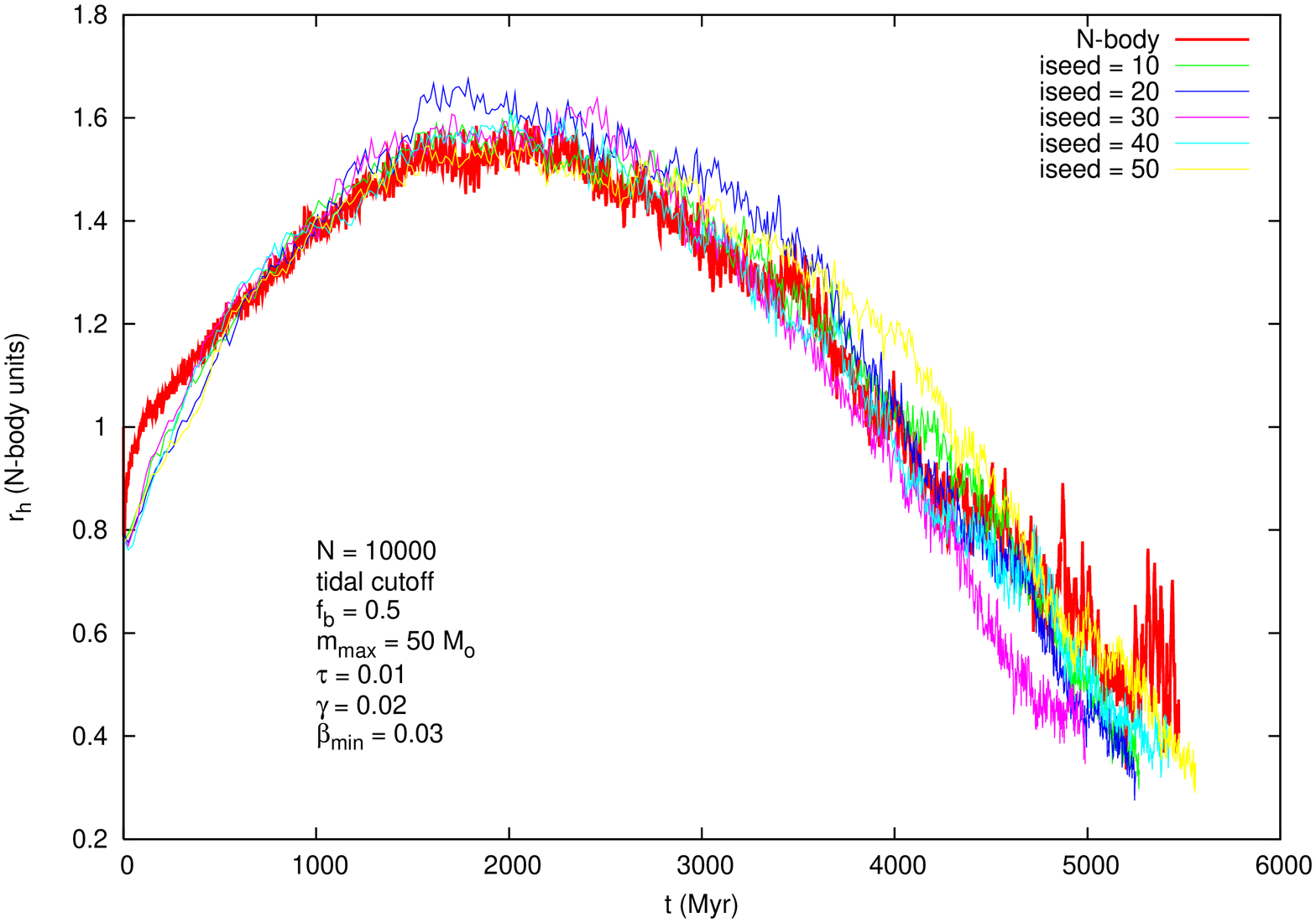}
\caption{Comparison of the evolution of the total mass and half-mass
radius in $N$-body and Monte Carlo simulations for models with tidal 
cutoff. The initial conditions are given in Tab. \ref{table:comparison} 
and described in the figures. The $N$-body model is the heavy continuous 
line, and the others are Monte-Carlo simulations with different initial 
random number sequence $iseed=10,20,30,40,50$.}
\label{fig:tc_iseed_comparison}
\end{figure}
As can be seen from Fig. \ref{fig:tc_iseed_comparison} the spread between 
models with exactly the same parameters, but with different initial 
random number sequence (iseed), is substantial, even for $N = 15000$. This spread is even 
larger for $N=2500$, as can be expected from theory. The spread 
between results with different $\beta_{min}$ and $\tau$ is well 
inside the spread connected with different iseed. Only the spread 
between models with different $\gamma$ is larger that the one connected 
with different iseed. The best values of the free code parameters are: 
$\gamma = 0.2$, $\beta = 0.03$ and $\tau = 0.01$.

\begin{figure}
\includegraphics[angle=-90, width=8cm]{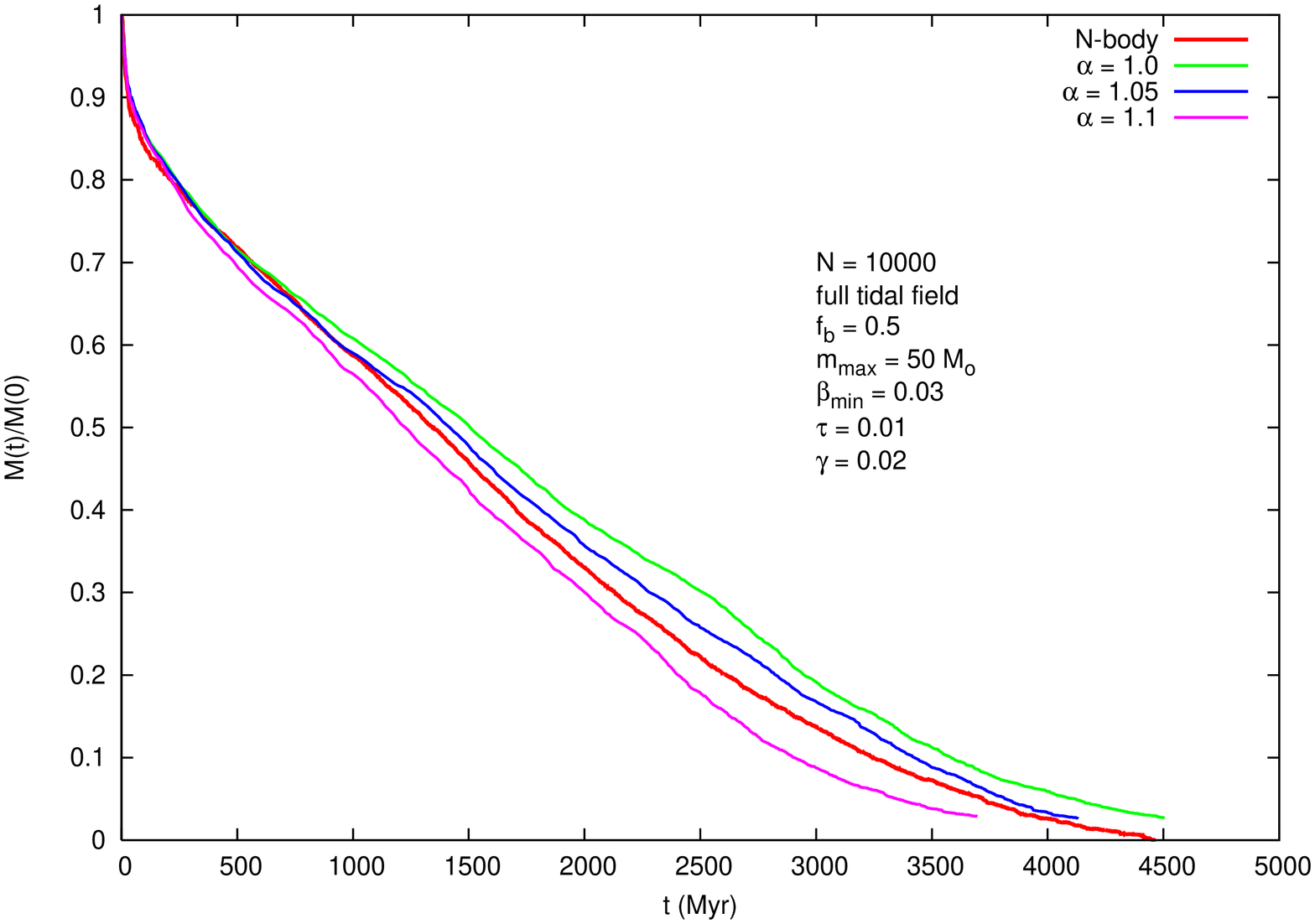}
\includegraphics[angle=-90, width=8cm]{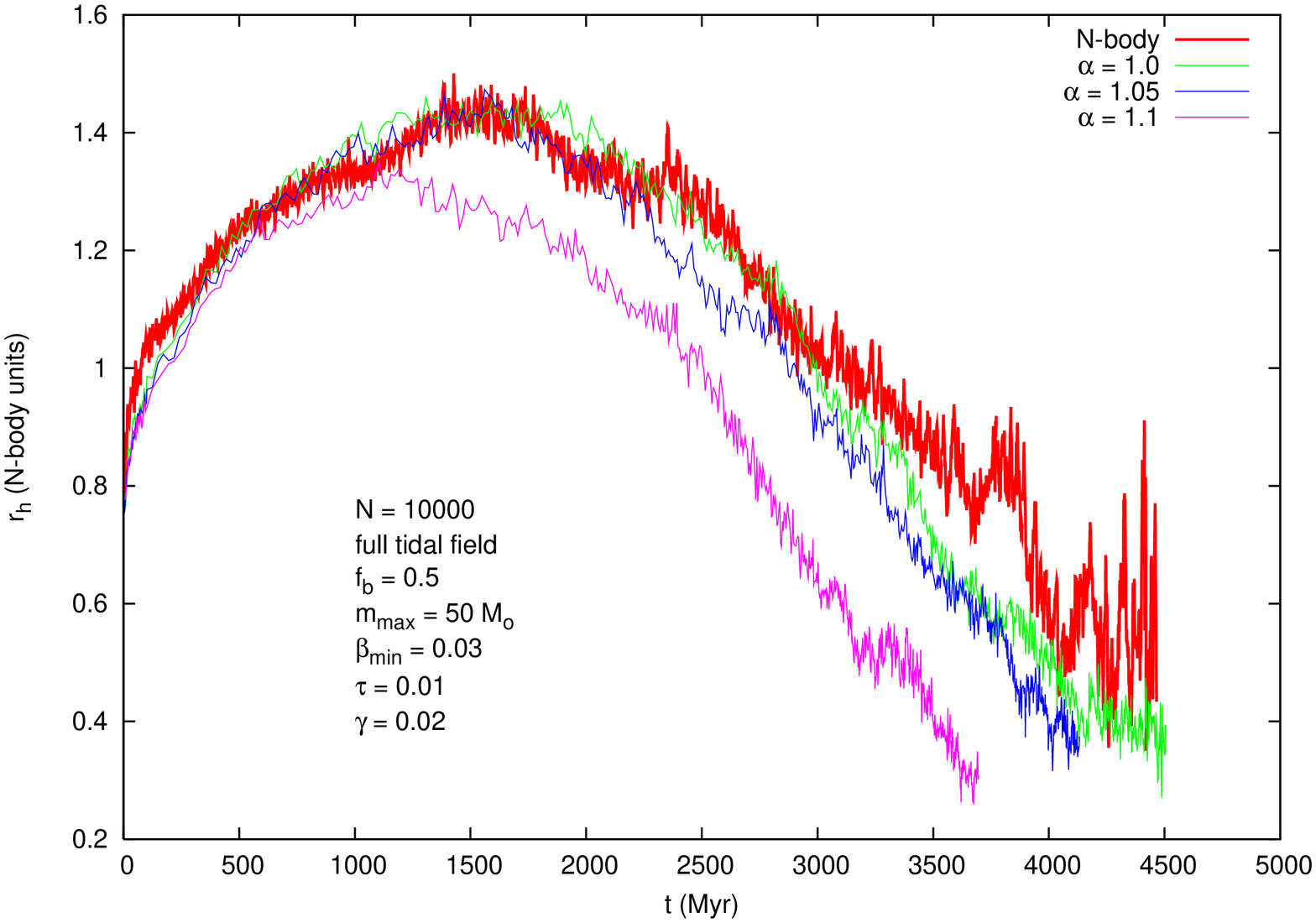}
\caption{Comparison of the evolution of the total mass and the 
half-mass radius in $N$-body and Monte Carlo simulations for models 
with full tidal field. The initial conditions are given in 
Tab. \ref{table:comparison} and described in the figures. 
The $N$-body model is the heavy continuous line, and the others 
are Monte-Carlo simulations with (from the right) $\alpha=1.0,1.05,1.1$.}
\label{fig:tf_alfa_comparison}
\end{figure}

\subsubsection{Models with full tidal field}\label{sec:2.2.2}

The process of escape from a cluster in a steady tidal field is extremely 
complicated. Some stars which fulfil the energy criterion (binding energy 
of the star greater than the critical energy $E_{t_f} =
-1.5(GM/r_{t})$, see Spitzer (1987)) 
can  still be trapped inside the potential well. Those stars can be scattered 
back to lower energy before they escape from the system. As  was pointed out 
by \citet{baum2001} these mechanisms  cause the cluster lifetime
to scale nonlinearly with relaxation time, in contrast
with what would be expected from the standard theory.
The efficiency of these effects  decreases as the 
number of stars increases. To account for this  in the Monte Carlo code an
additional free parameter was introduced according to the theory presented by
\citet{baum2001}. The critical energy for escaping stars was approximated by:
$E_{t_f} = -\alpha (GM/r_{t})$, where $\alpha = 1.5 - a (ln(\gamma N)/N)^{1/4}$. 
Thus the effective tidal radius for Monte Carlo simulations is
$r_{t_{eff}} = r_{t}/\alpha$ and it is smaller than $r_{t}$. This
leads to the result
that for Monte Carlo simulations a system is slightly too concentrated
compared
to $N$-body simulations, but the evolution of the total mass is well
reproduced, as well as the scaling of the dissolution time with $N$.

\begin{figure}
\includegraphics[angle=-90, width=8cm]{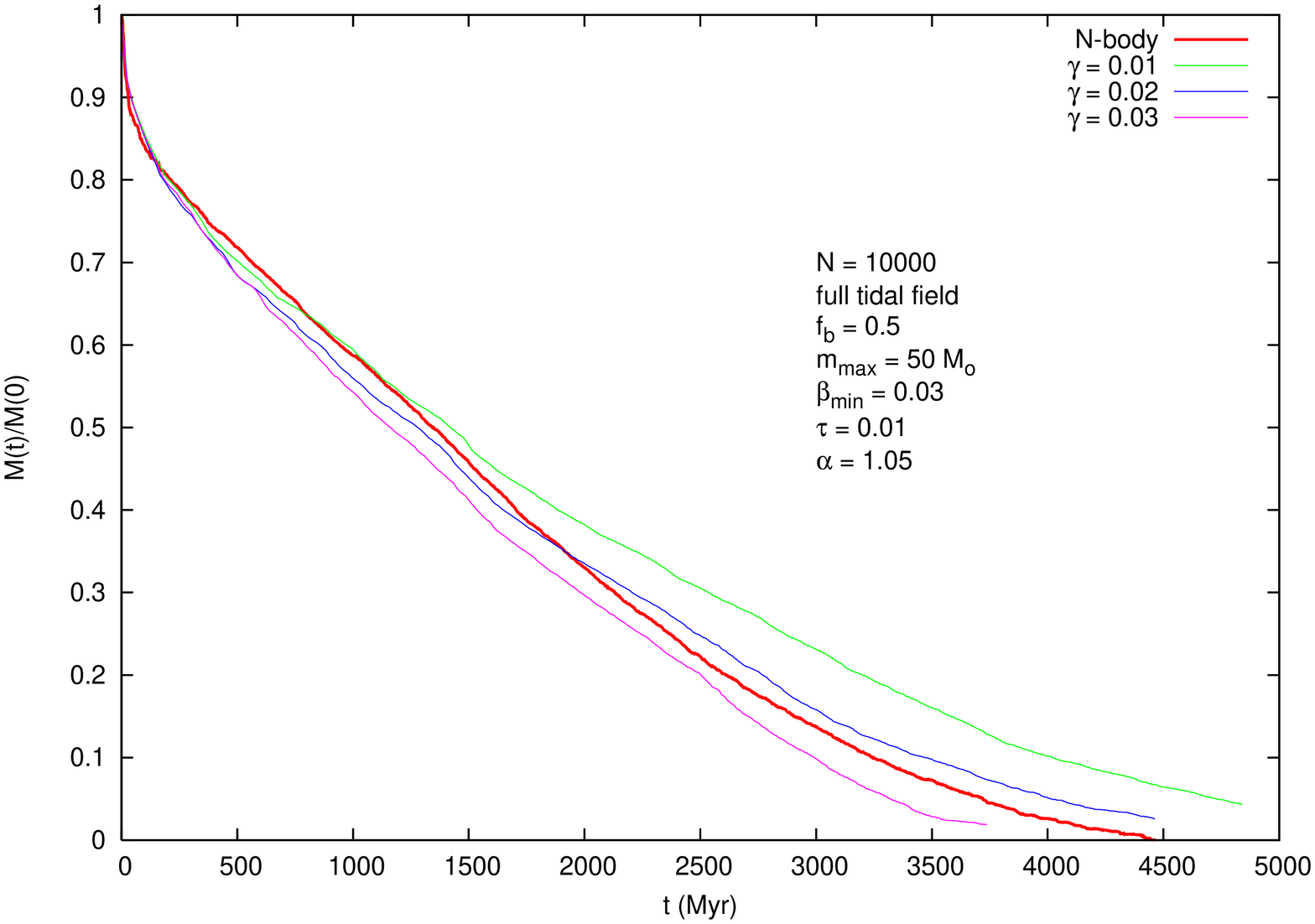}
\includegraphics[angle=-90, width=8cm]{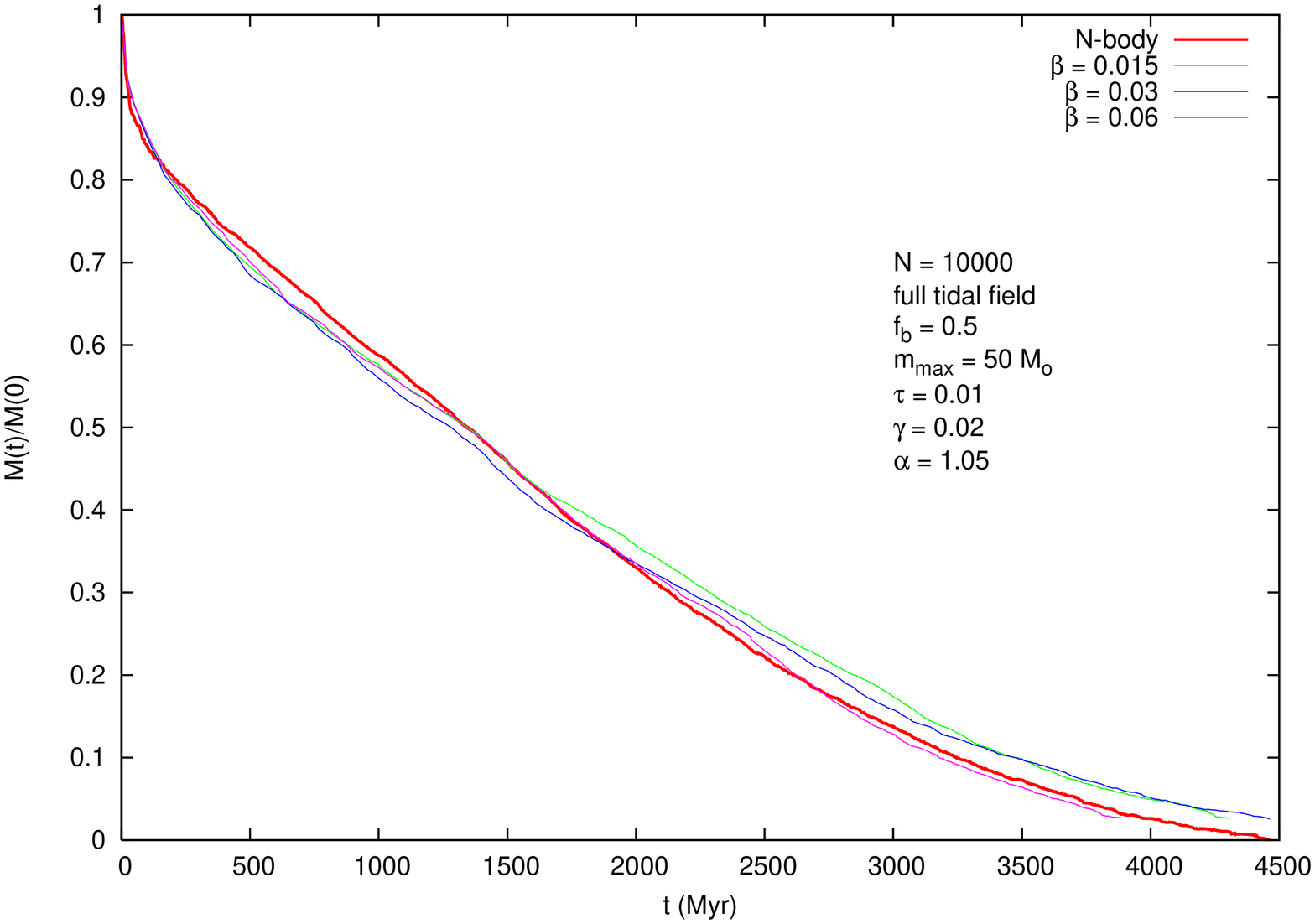}
\includegraphics[angle=-90, width=8cm]{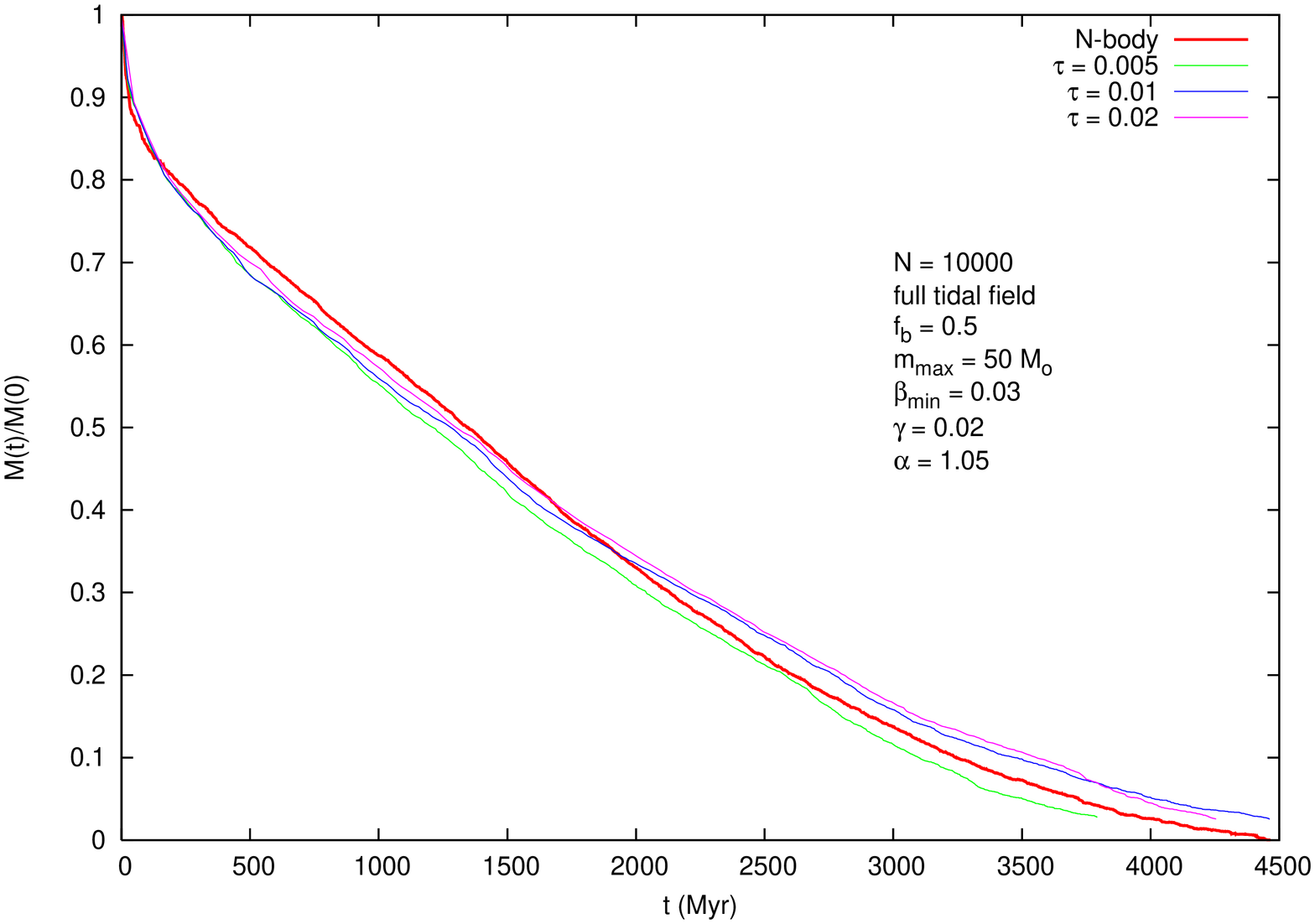}
\caption{Comparison of the evolution of the total mass in $N$-body and 
Monte Carlo simulations for models with full tidal field. The initial 
conditions are given in Tab. \ref{table:comparison} and described in the 
figures. The $N$-body model is the heavy continuous line, and the others 
are Monte-Carlo simulations with (from the left) $\gamma = 0.03, 0.02, 0.01$
(top),
(from the ) $\beta=0.06,0.03,0.015$ (middle), and 
 (from the right) $\tau=0.02,0.01,0.005$ 
(bottom).}
\label{fig:tf_gamma_beta_tau_comparison}
\end{figure}

 Fig. \ref{fig:tf_alfa_comparison}  shows the evolution of the total mass
and the half-mass radius
for different $\alpha$ for $N = 10000$. The other free parameters for the case 
of a full tidal field are the same as for the tidal cutoff case: $\gamma = 0.02$, 
$\tau = 0.01$ and $\beta_{min} = 0.03$. 
As can be seen by comparing   Fig.\ref{fig:tf_gamma_beta_tau_comparison} 
(lower two panels) with Fig.\ref{fig:tc_iseed_comparison} (top panel), 
again the spread between models with different $\beta_{min}$ and $\tau$ is well 
inside the spread connected with different iseed. The statistical spread also 
does not substantially interfere with the determination of
$\alpha$ and $\gamma$ (see Fig.\ref{fig:tf_gamma_beta_tau_comparison} (top panel) 
for $\gamma$).

\begin{figure}
\includegraphics[angle=-90, width=8cm]{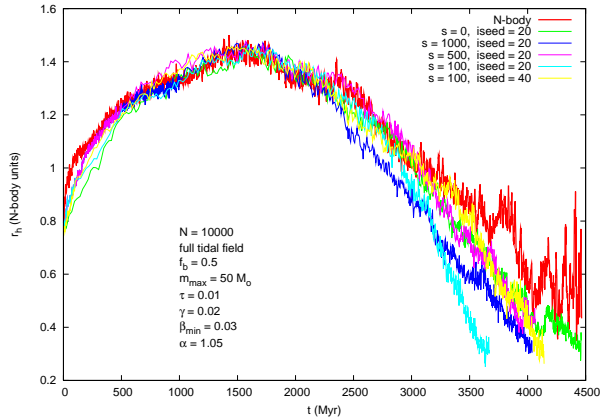}
\caption{Comparison of the evolution of the half-mass radius in 
$N$-body and Monte Carlo simulations for models with a full tidal field. 
The initial 
conditions are given in Tab. \ref{table:comparison} and described in the 
figures. The $N$-body model is the heavy continuous line, and the 
others are Monte-Carlo simulations with different time, $s$, up to
which the overall time step is half of the standard one. The model 
with $s=100$ has two different statistical realisations with
$iseed=20,40$. Note that some curves are overprinted (and hence
invisible) at early times.}
\label{fig:tf_iseed_s_comparison}
\end{figure}

As can be seen in Fig. \ref{fig:tf_alfa_comparison} (lower
panel) the evolution of $r_h$ 
up to time about 0.5 Gyr is slightly too slow in comparison to
$N$-body results. (This effect is even more pronounced for
smaller $N$: see Fig.\ref{fig:rh_tc_comparison}, top panel.)
This behaviour is connected with the way in which the effect of stellar 
mass loss is fed in to the cluster. In the Monte Carlo model the stellar 
evolution mass loss is postponed until the end of the overall time step, 
usually several Myr. So, for the most massive stars the stellar evolution 
can be substantially delayed and the cluster expands slower. 
In Fig. \ref{fig:tf_iseed_s_comparison} the evolution of $r_h$ is presented 
for different models in which  the overall 
time step was reduced by factor of two up to a certain time, $s$. 
It is clear that reduction of 
the overall time step in the phases of cluster evolution in which the most 
massive stars end their evolution helps to bring the Monte
Carlo results close to the
$N$-body ones. In later phases of cluster evolution, in which the time-scale 
of stellar evolution becomes larger than the half-mass relaxation time, 
the evolution does not depend systematically on the chosen overall time 
step. (In the simulations used in the determination of the free code parameters 
the adopted overall time step was a compromise between accuracy and speed.)

\begin{figure}
\includegraphics[angle=-90, width=8cm]{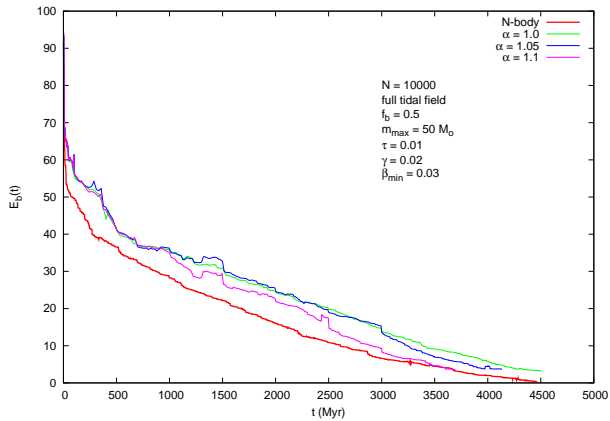}
\caption{Comparison of the evolution of the binary binding energy in
$N$-body and Monte Carlo simulations for models with full tidal field. 
The initial conditions are given in Tab. \ref{table:comparison} and 
described in the figures. The $N$-body model is the heavy continuous 
line, and the others are Monte-Carlo simulations  with
$\alpha=1.1,1.05,1.0$.}
\label{fig:tf_binen_comparison}
\end{figure}

\begin{figure}
\includegraphics[angle=-90, width=8cm]{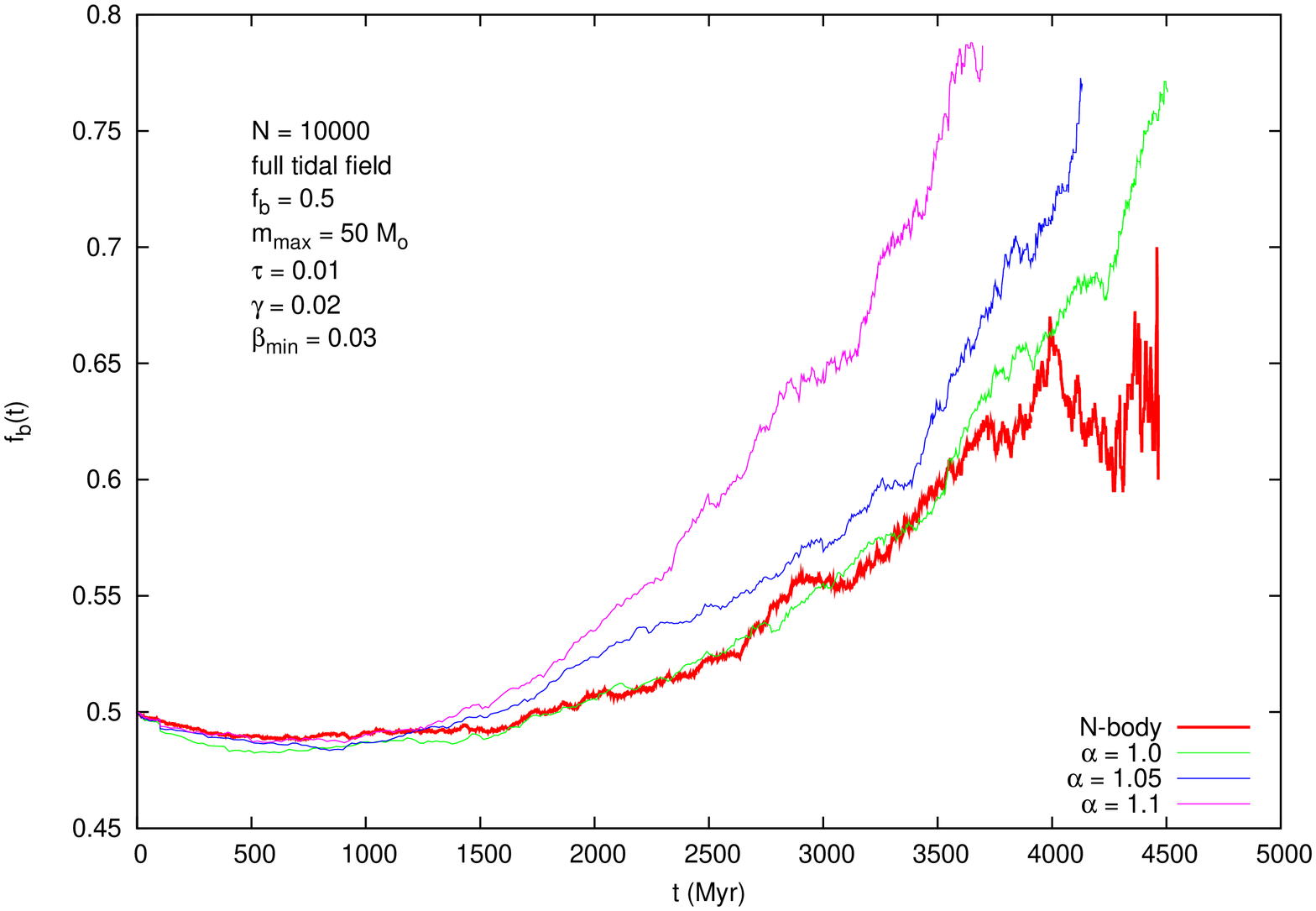}
\caption{Comparison of the evolution of the binary fraction in $N$-body 
and Monte Carlo simulations for models with full tidal field. The initial 
conditions are given in Tab. \ref{table:comparison} and described in the 
figures. The $N$-body model is the heavy continuous line, and the 
others are Monte-Carlo simulations  with $\alpha=1.1,1.05,1.0$.}  
\label{fig:tf_binfrac_comparison}
\end{figure}

\begin{figure}
\includegraphics[angle=-90, width=8cm]{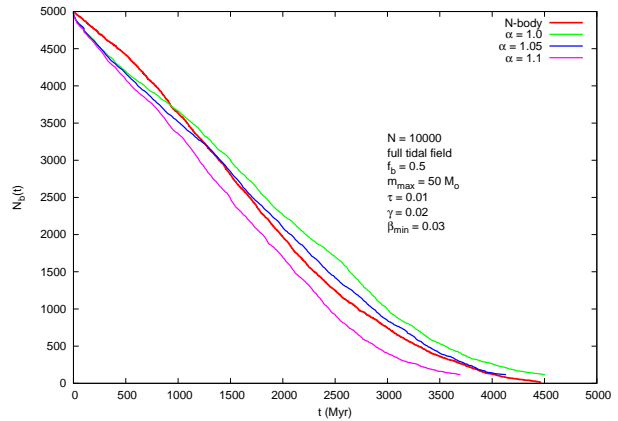}
\caption{Comparison of the evolution of the number of binaries in 
$N$-body and Monte Carlo simulations for models with full tidal 
field. The initial 
conditions are given in Tab. \ref{table:comparison} and described in the 
figures. The $N$-body model is the heavy continuous line, and the 
others are Monte-Carlo simulations  with $\alpha=1.1,1.05,1.0$.}
\label{fig:tf_binnum_comparison}
\end{figure}

As can be seen in Figs. \ref{fig:tf_binen_comparison}, 
\ref{fig:tf_binfrac_comparison}, \ref{fig:tf_binnum_comparison} the 
Monte Carlo code can reproduce
$N$-body simulations not only in respect of the global parameters of the
system, but also in respect of  properties connected with binary
activity. Despite the fact that the total number of binaries in the 
system and the binary fraction agree quite well  with $N$-body
simulations,
the total binding energy is substantially too high for the Monte Carlo 
simulations. This is connected with the fact that the present Monte Carlo 
simulations cannot follow  3- and 4-body interactions directly as the
$N$-body code does. Binaries can only harden or dissolve.
Therefore much of the
complexity of binary dynamical interactions is missing in the present 
Monte Carlo simulations. 

\begin{figure}
\includegraphics[angle=-90, width=8cm]{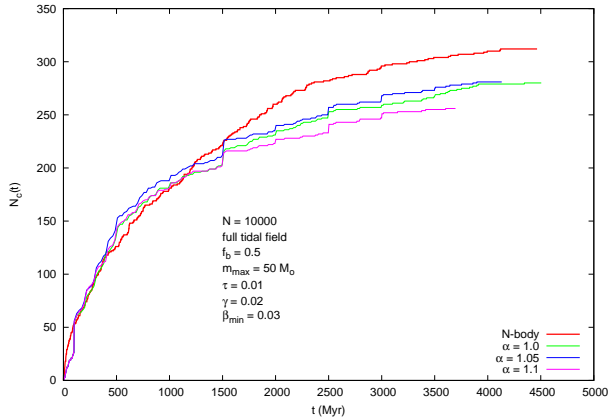}
\caption{Comparison of the evolution of the number of collisions in 
$N$-body and Monte Carlo simulations for models with full 
tidal field. The initial 
conditions are given in Tab. \ref{table:comparison} and described in the 
figures. The $N$-body model is the heavy continuous line, and the 
others are Monte-Carlo simulations  with $\alpha=1.1,1.05,1.0$.}  
\label{fig:tf-col_comparison}
\end{figure}

In Fig. \ref{fig:tf-col_comparison} is shown the number of collisions 
as a function of time for the Monte Carlo and $N$-body models. The 
collisions are mainly connected with binary mergers due to stellar 
evolution or dynamical binary interactions. There are only a few 
direct physical collisions between single stars. Up to time about 
1.5 Gyr  both models give similar results, and then the $N$-body 
model shows a larger number of collisions than the Monte Carlo model. 
Again this can be attributed to the fact that in the Monte Carlo 
simulations the complex dynamical binary interactions are not 
followed. In the Monte Carlo code a binary can only coallesce if,
{\sl after} the dynamical interaction, the periastron distance is smaller 
than the sum of the stellar radii. In the $N$-body code  binary 
coallescence occurs if, {\sl during} the
interaction, the distance between two stars 
is smaller than the sum of their radii. Definitely the latter can 
happen more frequently, in the case of strong interactions such
as prolonged resonances (temporary capture). When 
most collision events are connected with the stellar 
evolution of nearly isolated binaries, then both models agree.

\subsubsection{Model of M67}\label{sec:2.2.3}

The old open cluster M67 is an ideal testing ground for modelling the
interactions between dynamical and stellar evolution.  It has a
substantial population of blue stragglers, which are almost certainly
the product of stellar collisions, or mergers within primordial
binaries.  Also, it is small enough that its entire life history can
be modelled with $N$-body techniques \citep{hurleyetal2005}, though this has
become possible only within the last few years: though its present
mass (of order $2000\msun$) makes it seem an easy target for
simulation, its initial mass is likely to have been much higher (Tab. \ref{table:m67}).
This table also specifies the other initial parameters for our model,
which closely follow the  prescription of \citet{hurleyetal2005}.
The initial value of the tidal radius is determined by scaling the
value in their model at 4Gyr to the initial mass of our model.

\begin{table}
\caption{Initial conditions for M67}
\begin{tabular}{ll}
	$N_s+N_b$&$24000$\\
	M(0)&$1.904\times10^4M_\odot$\\
	Initial model&Plummer\\
	Initial tidal radius&$32.2$pc\\
	Initial mass function&\citet{kroupaetal1993}\\
	&with $\alpha_1 =1.3$\\
	IMF of binaries&\citet{kroupaetal1991}, eq.(1)\\
	Binary fraction&$N_s/(N_s+N_b) = 0.5$\\
	Binary eccentricities&$f(e) = 2e$\\
	Binary semi-major axes&Uniformly distributed in the
	logarithm\\
	& in the range $2(R_1+R_2)$ to $50$AU\\
	Run time (Monte Carlo)&7 min\\
	Run time (NBODY6)& 1 month
\end{tabular}
\label{table:m67}
\end{table}

The data from $N$-body simulations of M67 \citep{hurleyetal2005} were used to  
calibrate the last remaining parameter of the Monte Carlo code, namely $\alpha$. 
The inferred formula 
is $\alpha = 1.5 - 3.0 (ln(\gamma N)/N)^{1/4}$. The comparison of results 
from $N$-body and Monte Carlo simulations for M67 confirmed the values 
of $\gamma$, $\tau$ and $\beta_{min}$ found for smaller $N$ systems.

\begin{table}
\caption{Monte Carlo and $N$-body results for M67 at 4 Gyr}
\begin{tabular}{c|cc} \hline
 \qquad &$N$-body \citep{hurleyetal2005}& This work \\ \hline
$M/M_{\odot}$         & 2037 & 1984 \\
$f_b$                 & 0.60 & 0.59 \\
$r_t$ $pc^{-1}$       & 15.2 & 15.1 \\
$r_h$ $pc^{-1}$       & 3.80 & 3.03 \\
$M_L/M_{\odot}$       & 1488 & 1219 \\
$M_{L10}/M_{\odot}$   & 1342 & 1205 \\
$r_{h,L10}$ $pc^{-1}$ & 2.70 & 2.67 \\
\end{tabular}
\label{table:ourm67}
\medskip

L -- stars with mass above $0.5 M_{\odot}$ and burning nuclear fuel

L10 -- the same as L but for stars contained within 10 pc
\end{table}

\begin{figure}
\includegraphics[angle=-90, width=8cm]{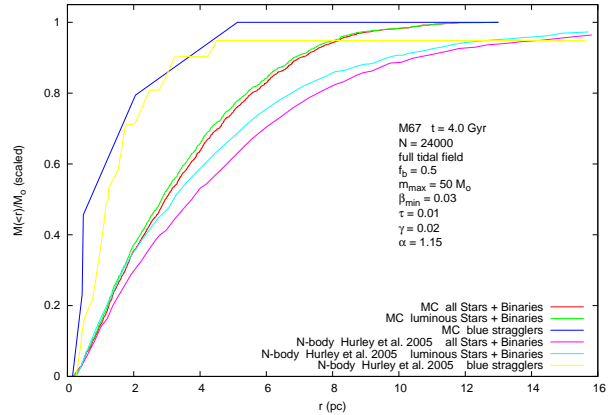}
\caption{Mass profiles of the Monte Carlo and $N$-body models of 
M67 at 4Gyr, scaled by the total mass involved in constructing the 
profile; from the right all mass, luminous mass (as defined in the text) 
and blue stragglers. The initial conditions are given in 
Tab. \ref{table:m67} and described in the 
figure.}
\label{fig:m67_mass_prof}
\end{figure}

The results of a comparison are summarised in Tab. \ref{table:ourm67}.
Taking into account the intrinsic statistical fluctuations of both methods
the results presented in Tab. \ref{table:ourm67}  show reasonably 
good agreement. At the time of 4 Gyr, when the comparison was done, 
both models consist of only a small fraction of the initial number 
of stars, making  fluctuations even more important. The values of the 
half-mass radius $r_h$ suggest that the Monte Carlo 
model is slightly too concentrated by comparison with the
$N$-body model. This, however, can be attributed to the treatment of the tide
(Sec.\ref{sec:2.2.2}), which leads to a smaller effective 
tidal radius than the tidal radius inferred from $N$-body simulations. 
Additionally, in $N$-body simulations, stars are considered
as escapers only
if their distance from the cluster centre is larger than $2r_{t}$ 
\citep{hurleyetal2005}. The mass outside  $r_{t}$ is  about 
$100 M_{\odot}$ (see Fig.\ref{fig:m67_mass_prof}), which is
small compared to the total cluster mass 
at any time, but nevertheless leads to slightly too large a
half-mass radius. Compared to $N$-body simulations, the
values shown by the Monte Carlo simulations for the luminous
mass $M_L$ and the
luminous mass inside 10 pc distance from the centre $M_{L10}$ are
too low. (The luminous 
mass is the mass of all stars with mass above $0.5 M_{\odot}$ and 
burning nuclear fuel \citep{hurleyetal2005}). The reasons for
this disagreement are unclear, but it is partly attributable to the smaller total
mass of the Monte Carlo model, and balanced by the somewhat larger
total mass in white dwarfs (see below).
Note, however, that the lower effective tidal radius, $r_{t_{eff}}$, 
in the Monte Carlo model forces  most stars 
to be confined inside 10 pc. Therefore $M_L$ and $M_{L10}$ are practically 
identical for the Monte Carlo simulations. Furthermore, the
mismatch between the two models is much smaller inside 10pc, which
suggests that the reason for the disagreement in $M_L$ concerns mainly large radii.

The spatial distributions of mass for the Monte Carlo and $N$-body models are 
illustrated in Fig. \ref{fig:m67_mass_prof}. There are only 4 blue stragglers 
in the model, compared with 20 in the $N$-body model. Agreement is not 
expected, because our model excludes one of the main channels for blue 
straggler production, i.e. collisions during resonant encounters. Despite 
the large difference in the number of blue stragglers present in 
both models, their mass distributions are very similar. Blue stragglers 
are more centrally concentrated then other luminous object in the cluster. 
The comparison with
the $N$-body model is qualitatively satisfactory, but quantitatively
reflects the larger half-mass radius and 
larger effective tidal radius in the $N$-body model. 

\begin{figure}
\includegraphics[angle=-90, width=8cm]{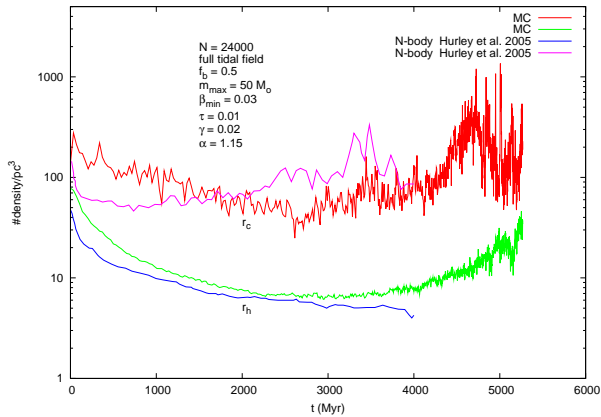}
\caption{Number density within the core and half-mass radii of the 
Monte Carlo and $N$-body models of M67, as a function of time. The 
initial conditions are given in Tab. \ref{table:m67} and described 
in the figure.}
\label{fig:m67_roc}
\end{figure}

Comparisons of the time-evolution of the number-density within the 
core and half-mass radii for Monte Carlo and $N$-body models are 
presented in Fig. \ref{fig:m67_roc}. Agreement between the two 
models is quite satisfactory. The observed difference can be
attributed to two factors: 
\begin{enumerate}
\item[(i)]  different definitions of the core radius are
used in the two models. 
In the Monte Carlo model the core radius is defined according to 
equation 1-34 in \citet{spitzer1987} and in the $N$-body model
from a density-weighted radius \citep{casertanohut1985};
\item[(ii)]  the Monte Carlo model is more centrally 
concentrated than the $N$-body model because of the smaller effective 
tidal radius $r_{t_{eff}}$. 
  
\end{enumerate}

\begin{figure}
\includegraphics[angle=-90, width=8cm]{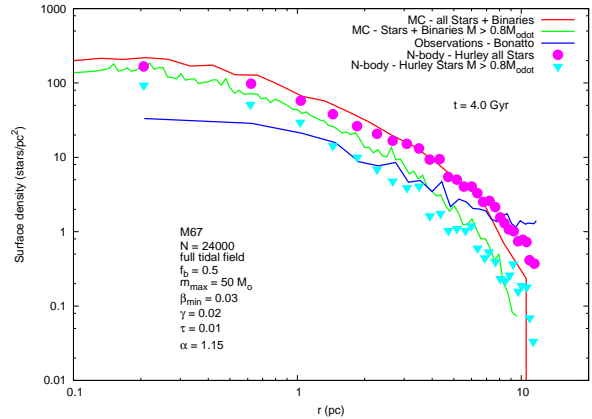}
\caption{Profile of the projected number density of all luminous stars and 
binaries and luminous stars above $0.8M_\odot$, compared with the observations 
of Bonatto \& Bica (extracted from \citet*[Fig.7]{hurleyetal2005}) and 
$N$-body simulations at 4 Gyr. The initial conditions are given in 
Tab. \ref{table:m67} and described in the figure.}
\label{fig:m67_sd}
\end{figure}

A comparison of the projected number density of luminous stars above
$0.8M_\odot$ for the Monte Carlo and $N$-body simulations
and from observations { (\citet{bb2005} as quoted in 
\citet{hurleyetal2005})} is presented in Fig. \ref{fig:m67_sd}. This
confirms the conclusions reached so far: the overall density of the
Monte Carlo model is slightly larger than that of the $N$-body model and the 
half-mass radius of the Monte Carlo model is too small. Both models exceed
the observed surface 
density in the central part of the system but underpredict it in 
the outer halo. These regions require separate discussion:
\begin{enumerate}
\item The higher observational value at large radii suggests contamination 
of the observed field by stars which are not members of the
cluster. Indeed, the data quoted in \citet{hurleyetal2005} are not 
corrected for the background density of stars. We also have not corrected 
the data for the background, in order to
analyse the simulation data in the same way as  in \citet{hurleyetal2005}
(see Fig. 7 there). (We consider the effect of the background in 
Sec.\ref{sec:refinement}).
\item It is possible that the observational value at small radii
could be lowered if it were supposed that the observations 
are not fully corrected for the large binary fraction in the core. 
Furthermore, as can be seen from the numerical simulations, the surface 
density is lower if only luminous stars are taken into account. So, 
corrections for a low-luminosity cutoff and unresolved binaries can help 
to bring observation and simulation closer.   
However, we should not expect a large correction factor
for contamination, because of the high latitude of M67.
\end{enumerate}
This discussion of the centre of M67 suggests that probably only some 
changes in the initial model of M67 can bring both 
observations and simulations into agreement, and we consider this in
Sec.\ref{sec:refinement}. (That was not our intention in the present
section, where our aim is to check that the Monte Carlo code can
produce results consistent with the $N$-body model for a realistic
cluster model.)

\begin{figure}
\includegraphics[angle=-90, width=8cm]{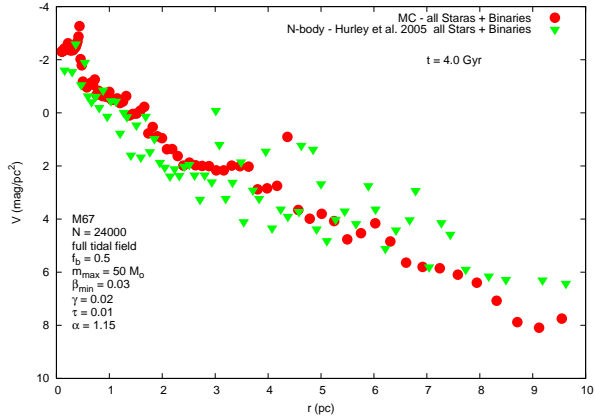}
\caption{Surface brightness profile of all luminous stars and binaries for
the Monte Carlo and $N$-body simulations at 4 Gyr. The initial conditions 
are given in Tab. \ref{table:m67} and described in the figure.}
\label{fig:m67_sb}
\end{figure}

In Fig. \ref{fig:m67_sb} the surface brightness profiles for the Monte Carlo and
$N$-body models are presented. To construct these surface brightness profiles 
all stars and binaries were used. The data are very noisy, particularly for the
$N$-body simulation. The agreement between the two models is reasonably good.
Again the conclusion reached before are confirmed. The surface brightness in the 
central parts of the system is slightly larger for the Monte Carlo model than 
that of $N$-body model and outside in the cluster halo the surface brightness
is larger for the $N$-body model. The latter is again connected with the effective
tidal radius for the Monte Carlo code which is smaller than the
nominal tidal radius for the two
models. Its effects are particularly clear in Fig.\ref{fig:m67_sd}.

\begin{figure}
\includegraphics[angle=-90, width=8cm]{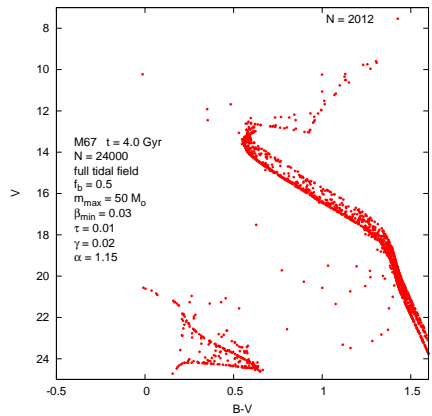}
\caption{Colour-magnitude diagram at 4 Gyr for Monte Carlo simulation. 
The initial conditions are given in Tab. \ref{table:m67} and described 
in the figure.}
\label{fig:m67_cmd}
\end{figure}

A form of colour-magnitude diagram is shown in Fig. \ref{fig:m67_cmd},
which can be compared with Fig.10 of \citet{hurleyetal2005}. The
resemblance is qualitatively satisfactory, except for the relative
paucity, already referred to, of blue stragglers in the Monte Carlo
model, and the shortness of the sequence of blue stragglers, compared to the
$N$-body model.

\begin{figure}
\includegraphics[angle=-90, width=8cm]{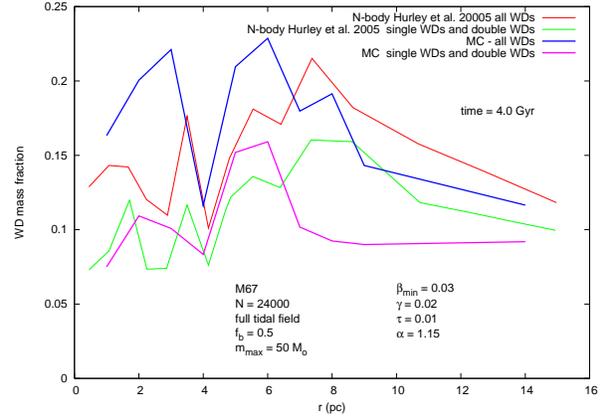}
\caption{Mass fraction of white dwarfs as a function of radius for 
Monte Carlo and $N$-body models at 4 Gyr. The results are for all 
WDs and only single WDs and double WDs. The initial conditions are 
given in Tab. \ref{table:m67} and described in the figure.}
\label{fig:m67_wd}
\end{figure}

\citet{hurleyetal2005} discuss the different exotic populations of
their model at some length. As already stated in connection with the
blue straggler population, however, our model lacks important
processes for the formation of such objects.  Therefore we confine
attention to more normal populations, namely white dwarfs.
The mass fraction of white dwarfs is $0.18$, slightly larger
than the value of $0.15$ given by \citet{hurleyetal2005}. 
Therefore the total mass in white dwarfs is larger in the Monte
Carlo model by about 60$\msun$. In
particular the white dwarf fraction in the central part of the 
system is larger for the Monte Carlo model than for the $N$-body model 
(Fig. \ref{fig:m67_wd}). The spatial distributions of white dwarfs
are similar in the two models. The maximum lies around 6 -- 8 pc,
and is more pronounced in the $N$-body model; in the Monte
Carlo model the profile could be flat (within fluctuations) below
this range of radii. The half-mass 
radius of the white dwarfs is 2.73 pc, much bigger than the value
of 0.6 pc reported by \citet{hurleyetal2005}; we believe their value may 
be in error.

\begin{figure}
\includegraphics[angle=-90, width=8cm]{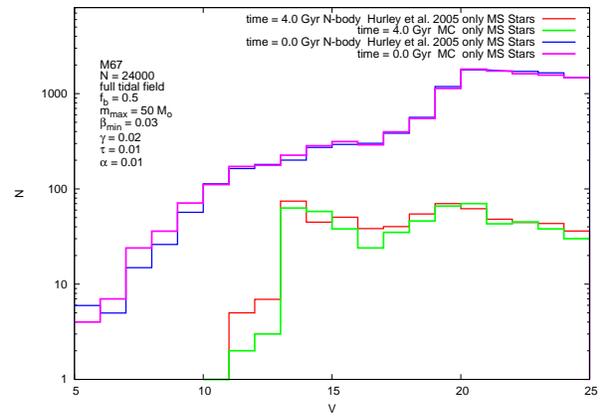}
\caption{Luminosity function of single main sequence stars for 
Monte Carlo and $N$-body models at 0 Gyr and 4 Gyr. The initial 
conditions are given in Tab. \ref{table:m67} and described in the figure.}
\label{fig:m67_lum}
\end{figure}

For single main sequence stars we present in Fig. \ref{fig:m67_lum} the 
luminosity functions for the Monte Carlo and $N$-body
models for times  0 Gyr 
and 4 Gyr. The luminosity functions for time 0 Gyr agree 
quite well for the two models. Only for very bright stars, $V < 10$ mag,
can one observe noticeable differences. They can be attributed to statistical
fluctuations connected with different realisations of the initial model.
For 4 Gyr our Monte Carlo result misses significant
numbers of stars at the high-luminosity end of the distribution,
which are found in the $N$-body simulation and which \citet{hurleyetal2005}
attribute to collisions; as already stated in
our discussion of blue stragglers, important channels for the
formation of such stars are missing at present in our model. For the low
mass end of the luminosity function the two models agree reasonably well.

\begin{figure}
\includegraphics[angle=-90, width=8cm]{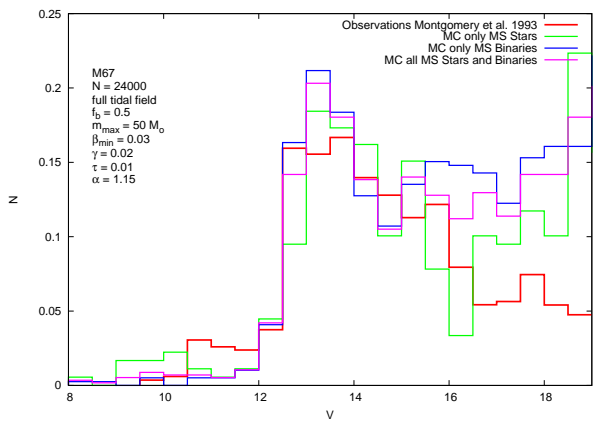}
\caption{Comparison of the luminosity function for the Monte Carlo model at 4 Gyr 
and observations (Montgomery at al. 1993), for main sequence 
stars only, binaries only and both main sequence stars and binaries. 
The initial conditions are given in Tab. \ref{table:m67} and described 
in the figure. { Note that all curves are normalised separately.}}
\label{fig:m67_lum_obs}
\end{figure}

Finally, the comparison between observations \citep{montgomeryetal1993} and 
the Monte Carlo simulations for the luminosity function is presented in
Fig. \ref{fig:m67_lum_obs}. The luminosity function was normalised by the 
number of stars with luminosity  $V < 15.5$ mag. In the figure are shown 
the luminosity functions for only main sequence stars, only main sequence 
binaries, and  all main sequence stars and binaries (for the Monte Carlo model) and
for main sequence stars and binaries (observations). It is clear that
the simulations do not agree with the observations. The luminosity function
for the Monte Carlo model is too low at the high-luminosity end, too high at the
low-luminosity end and too high near the main sequence turn-off ($V=13$ mag).
The drop in the luminosity function of the Monte Carlo
model at the high-luminosity end can be attributed to an underproduction
of blue stragglers in the Monte Carlo model comparable to observations. 
The excessive luminosity function at the low-luminosity
end and  around the main sequence turn off cannot be so easily 
explained. Even though observational issues may be relevant at the
faint end (Sec.\ref{sec:refinement}), some of these mismatches 
suggest that the initial model of M67 is wrong 
and some refinement is needed. Because the Monte Carlo model
agrees so well with the $N$-body model (Fig.\ref{fig:m67_lum}),
similar conclusions may be reached for a comparison of the $N$-body
model with observed luminosity function, except for the region
occupied by blue stragglers.

In this section it has been shown that the Monte Carlo code is able to 
follow the evolution of a realistic star cluster model at
a similar level of complexity to that of the 
$N$-body code. The data provided by the code is as detailed as for the 
$N$-body code and can be used for comparison with many kinds of observational data.
The next section will be devoted to further Monte Carlo modeling of the old open cluster
M67 in order to improve the match with its observational
properties and find possible initial conditions for this cluster.

\section{Refinement of the M67 model}\label{sec:refinement}

As was shown in the previous subsection (see Sec. \ref{sec:2.2.3}) the 
model of M67 proposed by
\citet{hurleyetal2005}, whether one uses an $N$-body code or a
Monte Carlo code, shows significant disagreement with such
observational properties of the cluster as the luminosity function and 
the surface density profile. Other observational data
about M67 are summarised in \citet{hurleyetal2005} and listed in
Tab. \ref{table:obs_m67}. Generally, however, the structural cluster parameters 
are not well known, and have been derived from more basic
data by some model-dependent analysis.  Therefore we prefer to
compare with the observational data as directly as possible.

\begin{table}
\caption{Properties of M67\ $^a$}
\begin{tabular}{ll} \hline
Distance from Sun         & 870 pc\\
Absolute distance modulus\ $^b$ & 9.44 \\
Distance from GC          & 6.8 -- 9.1 kpc\\
Orbit eccentricity        & 0.14\\
Luminous mass             & $\sim 1000 M_{\odot}$\\
Core radius               & 1.2 pc\\
Tidal radius              & $> 11.4$ pc\\
Half-mass radius\ $^c$    & $\sim 3.0$ pc\\
Binary fraction           & $\sim 50\%$\\
Z                         & $\sim  0.0$\\
Age                       & $\sim  4$ Gyr\\
$A_V = 3.25 E(B-V)$       & 0.16\\
\end{tabular}
\label{table:obs_m67}
\medskip

$^a$ References are given in \citet{hurleyetal2005}   

$^b$ \citep{montgomeryetal1993}; used in the calculation of the
luminosity function of of the model

$^c$ For main sequence stars with masses $\geq 0.87 M_{\odot}$ and 
within 10 pc
\end{table}

The observational data we shall use for comparison with the results of 
Monte Carlo simulations are: (i) the luminosity function -
\citep{montgomeryetal1993}, and
(ii) the surface density profile - \citep{bb2003,bb2005}. 
Unfortunately, the observational data for the surface density
profile also seem very uncertain, even though they were collected by 2MASS 
(Two Micron All Sky Survey); they differ by a factor larger than 1.5.
In all figures in which the surface density 
profiles will be presented there are two observational curves: 
(i) -  \citep{bb2005} (bottom), corrected for the background density at the
level of $0.73$ stars arcmin$^{-2}$, which was estimated at the distance 
about $25$ arcmin, and (ii) - \citep{bb2003} (top), corrected for the 
background density at the level of $4$ stars pc$^{-2}$, which was estimated 
at the limiting radius about $9$ pc.

The model of M67 presented in the previous section had three
problems: (i) it was too dense, (ii)  it
produced too flat a luminosity function for dim stars,
and (iii) it contained too 
many stars around the main sequence turn off. If we take
the observations at face value, to bring the model into better agreement
with observation it has to either lose more of its less massive
stars, or else contain
smaller numbers of those stars initially. Additionally, it has
to have some property such as
initial mass segregation or stronger energy generation in order to show 
a smaller concentration at the present day. We now
describe the parameter space which we explored in order to improve
the model.

The free parameters of the initial models are: (i) \  $N$, number of objects 
(stars and binaries), which we explored in the range between 22000 and 40000, 
(ii) \  $f_b$, binary fraction, in the range between 0.4 and 0.7,
(iii) \  $r_t$, tidal radius, between 30 and 38 pc, (iv) \  $r_t/r_h$,
i.e. the ratio between the
tidal and half-mass radii,  between 6 and 12, and (v) \ 
$\alpha_{IMF}$, the
power-law index of the low-mass part of the initial mass function (IMF), 
between 0.1 and 1.3. (The canonical value is $\alpha_{IMF} = 1.3$
\citep{kroupa2007}). 
Over 135 models of the old open cluster M67 were run. We did not
carry out this exploration very
systematically; rather on the basis of inspection of the 
results, the parameters of the next set of new models were chosen. 

\begin{figure}
\includegraphics[angle=-90, width=8cm]{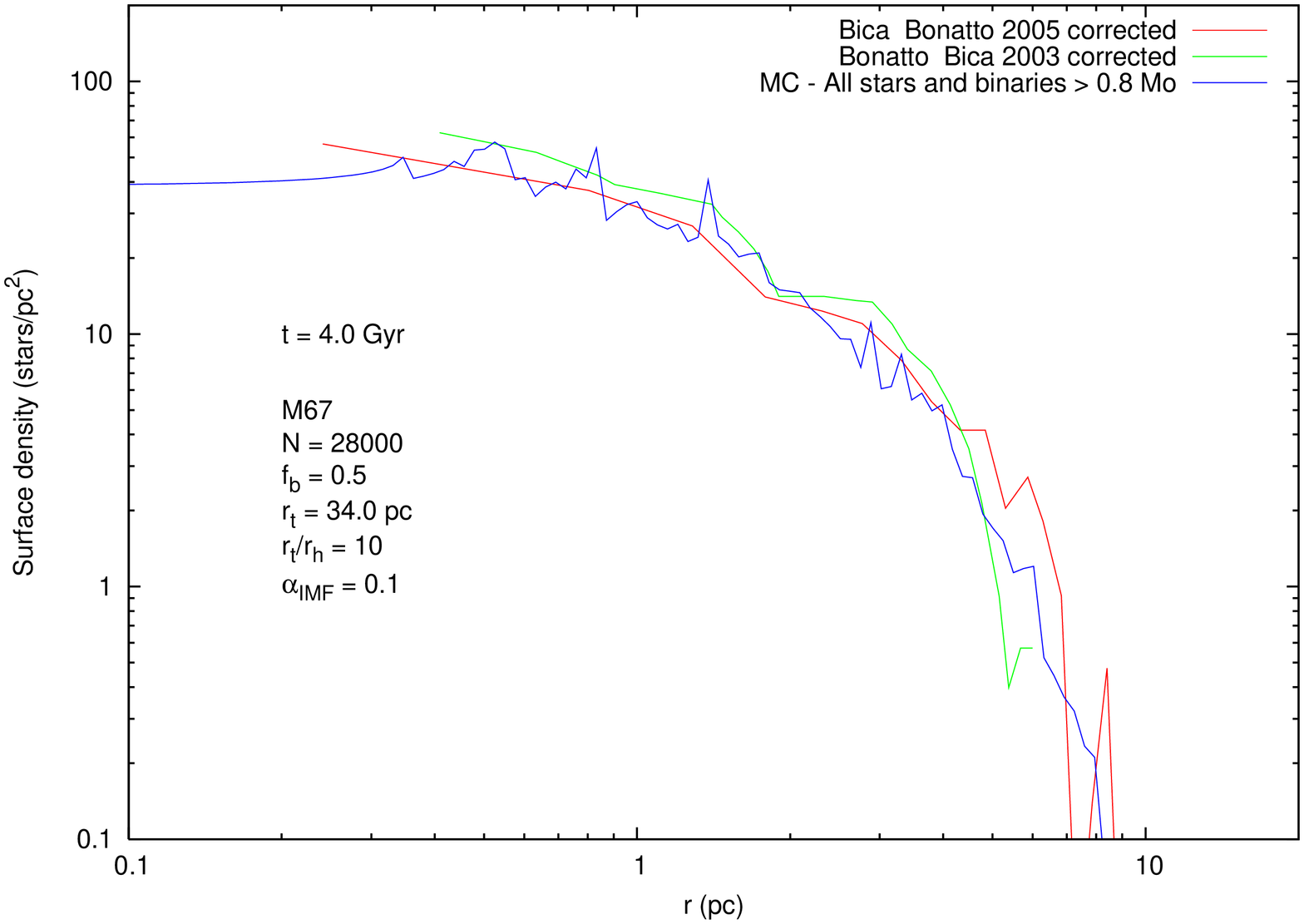}
\includegraphics[angle=-90, width=8cm]{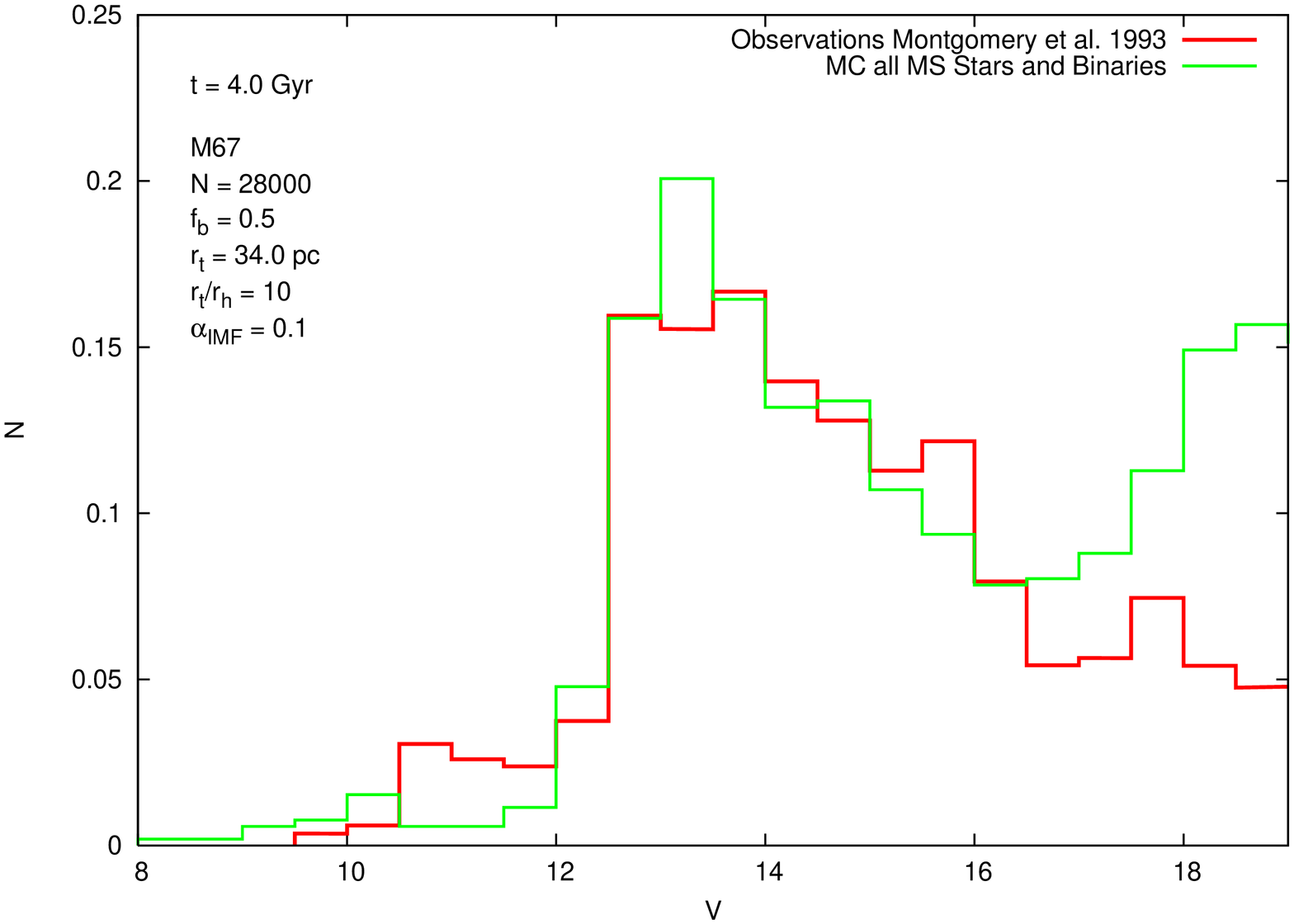}
\caption{Comparison of the surface density profile and luminosity function
for the Monte Carlo model for $\alpha_{IMF} = 0.1$ and observations. The initial 
model parameters are described in the figures. The observational data is 
described in the text}
\label{fig:m67_2000.1}
\end{figure}

\begin{figure}
\includegraphics[angle=-90, width=8cm]{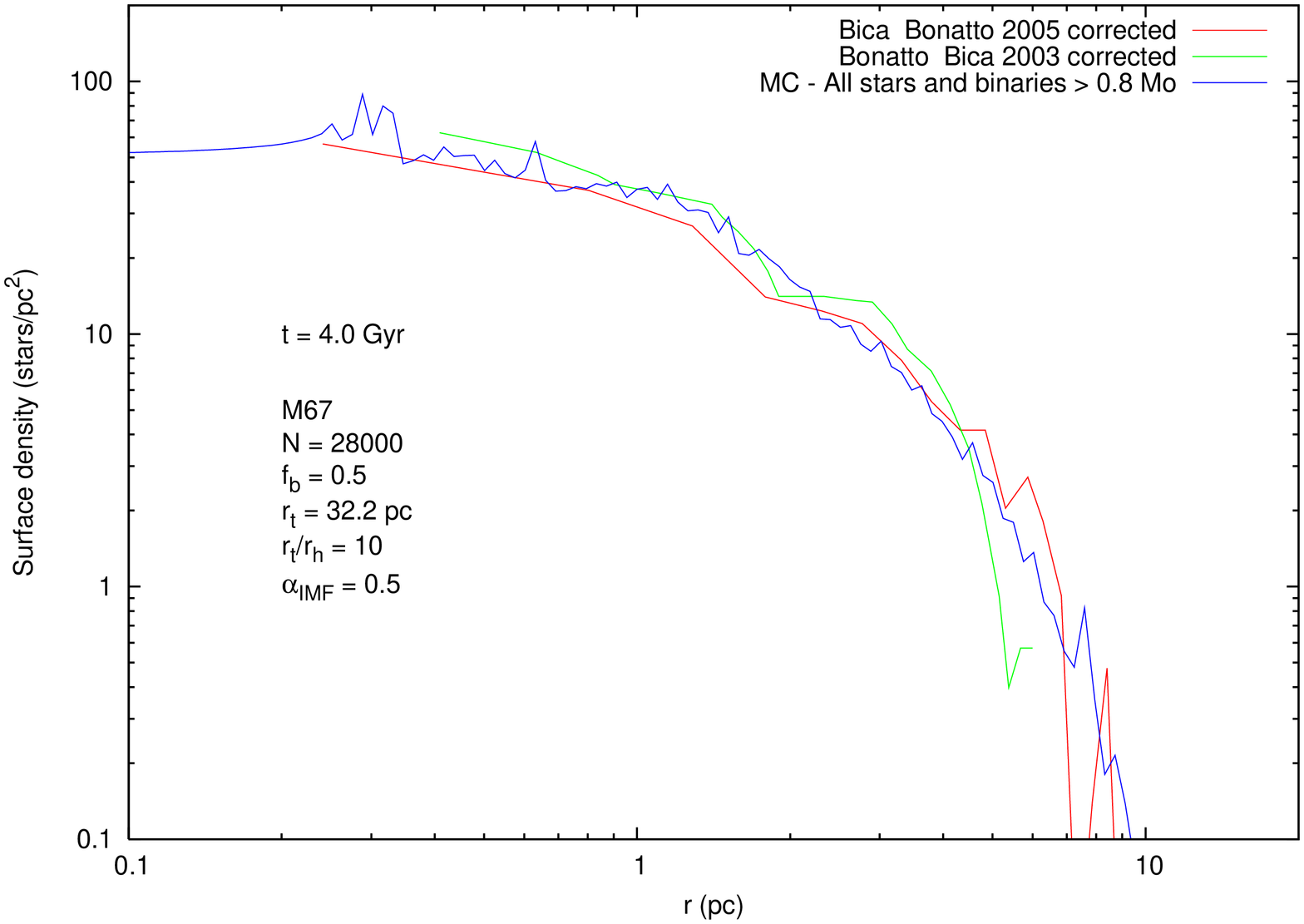}
\includegraphics[angle=-90, width=8cm]{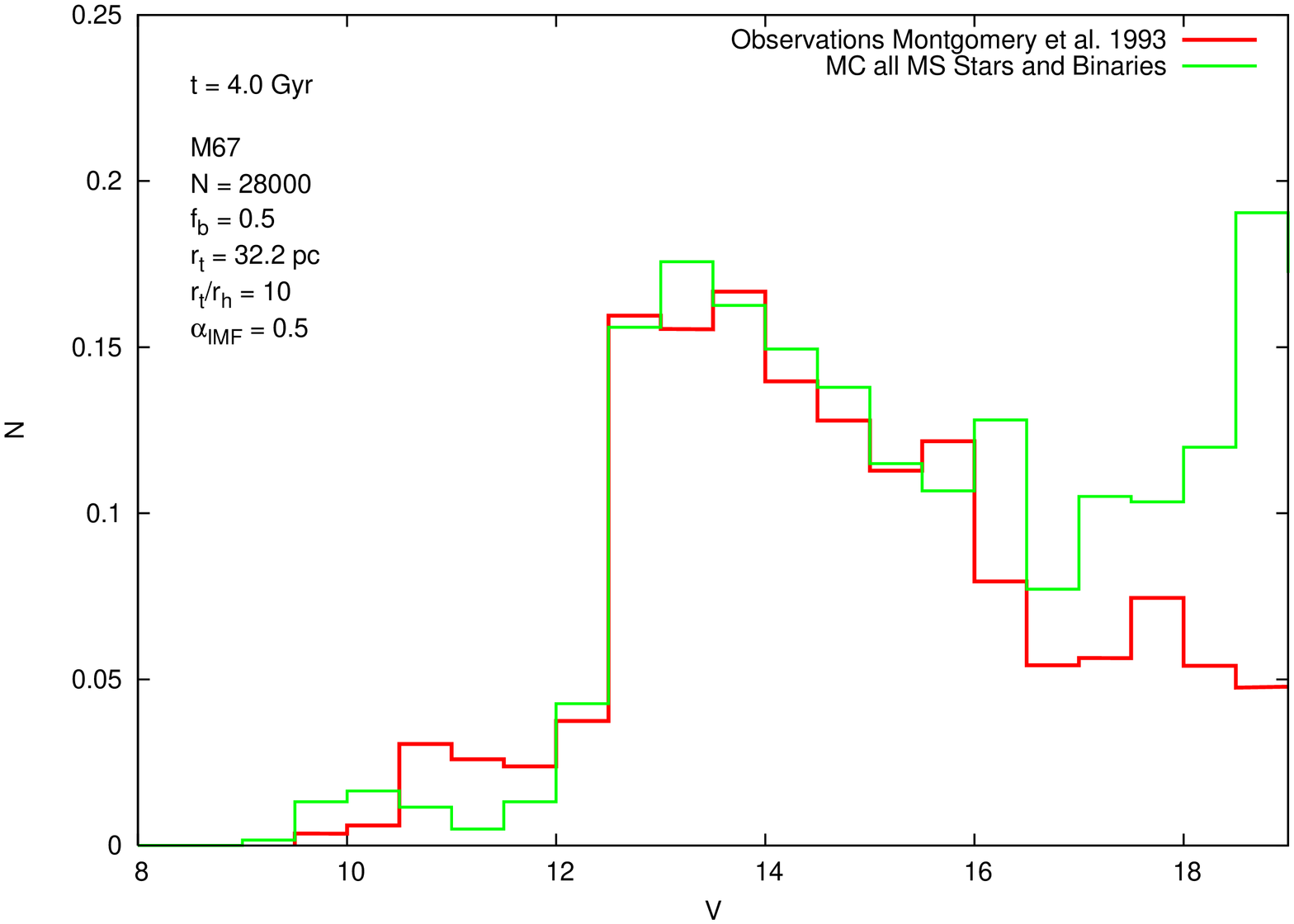}
\caption{Comparison of the surface density profile and luminosity function
for the Monte Carlo model for $\alpha_{IMF} = 0.5$ and observations. The initial 
model parameters are described in the figures. The observational data is 
described in the text}
\label{fig:m67_2000.5}
\end{figure}

\begin{figure}
\includegraphics[angle=-90, width=8cm]{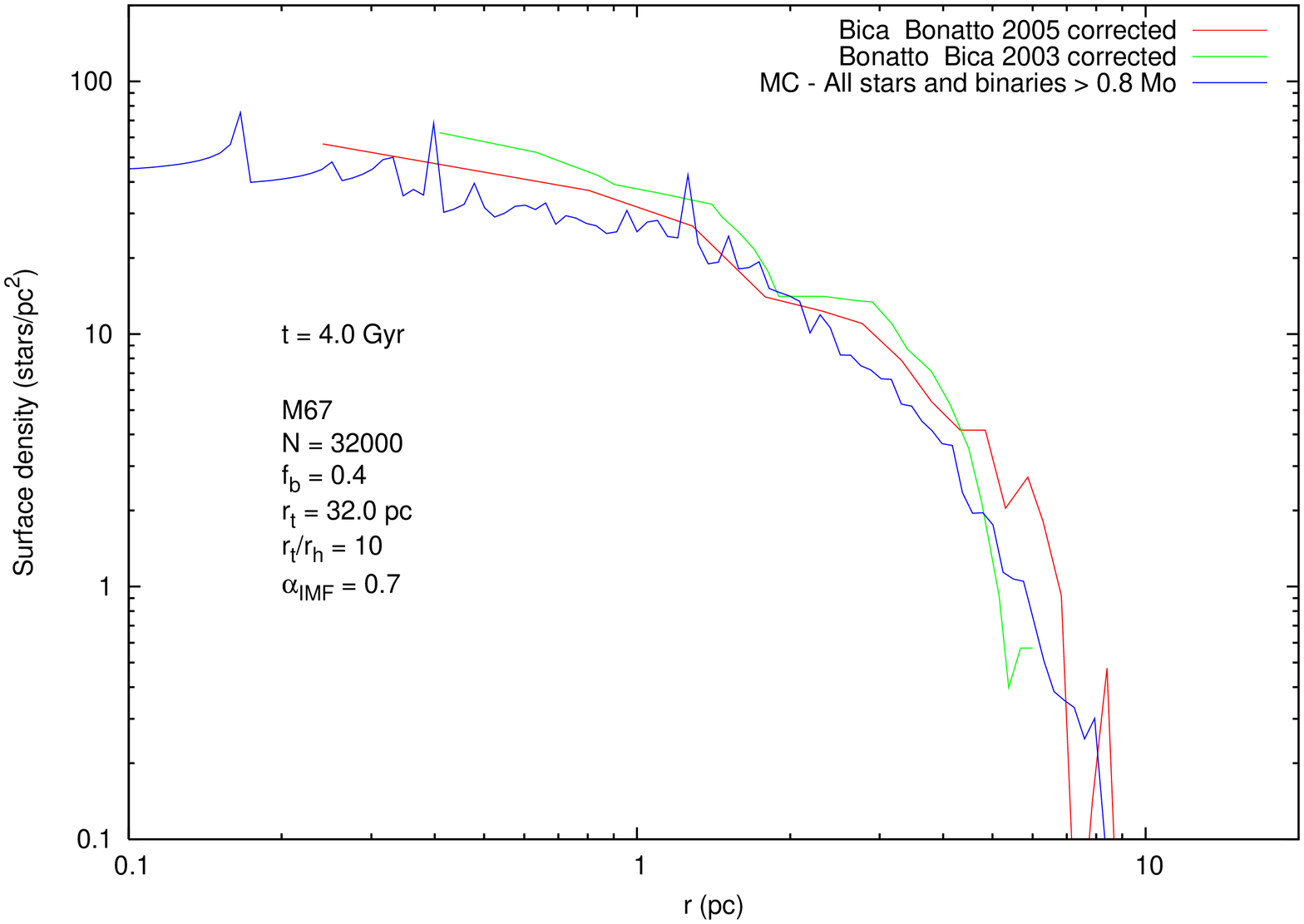}
\includegraphics[angle=-90, width=8cm]{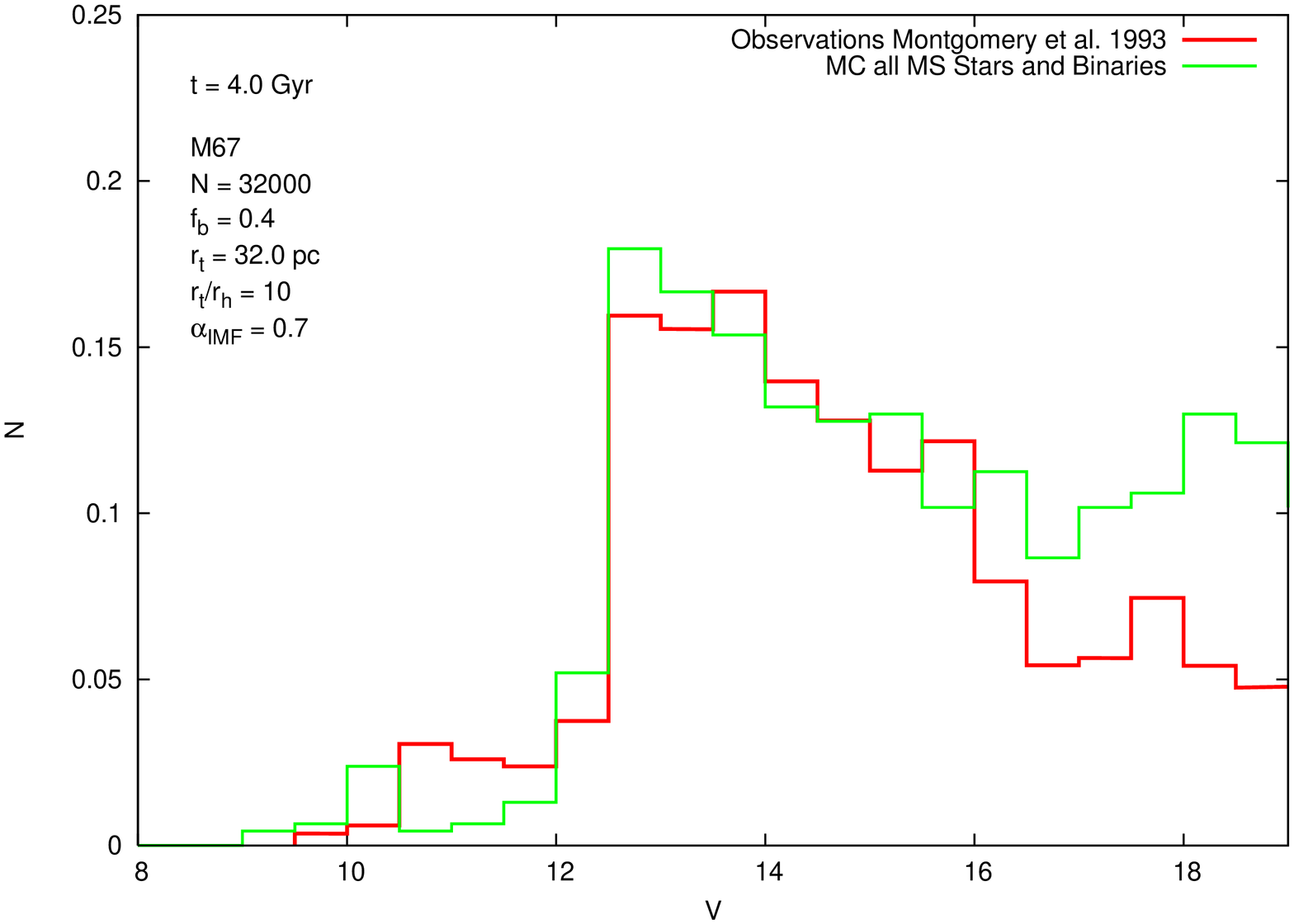}
\caption{Comparison of the surface density profile and luminosity function
for the Monte Carlo model for $\alpha_{IMF} = 0.7$ and observations. The initial 
model parameters are described in the figures. The observational data is 
described in the text}
\label{fig:m67_2000.7}
\end{figure}

\begin{figure}
\includegraphics[angle=-90, width=8cm]{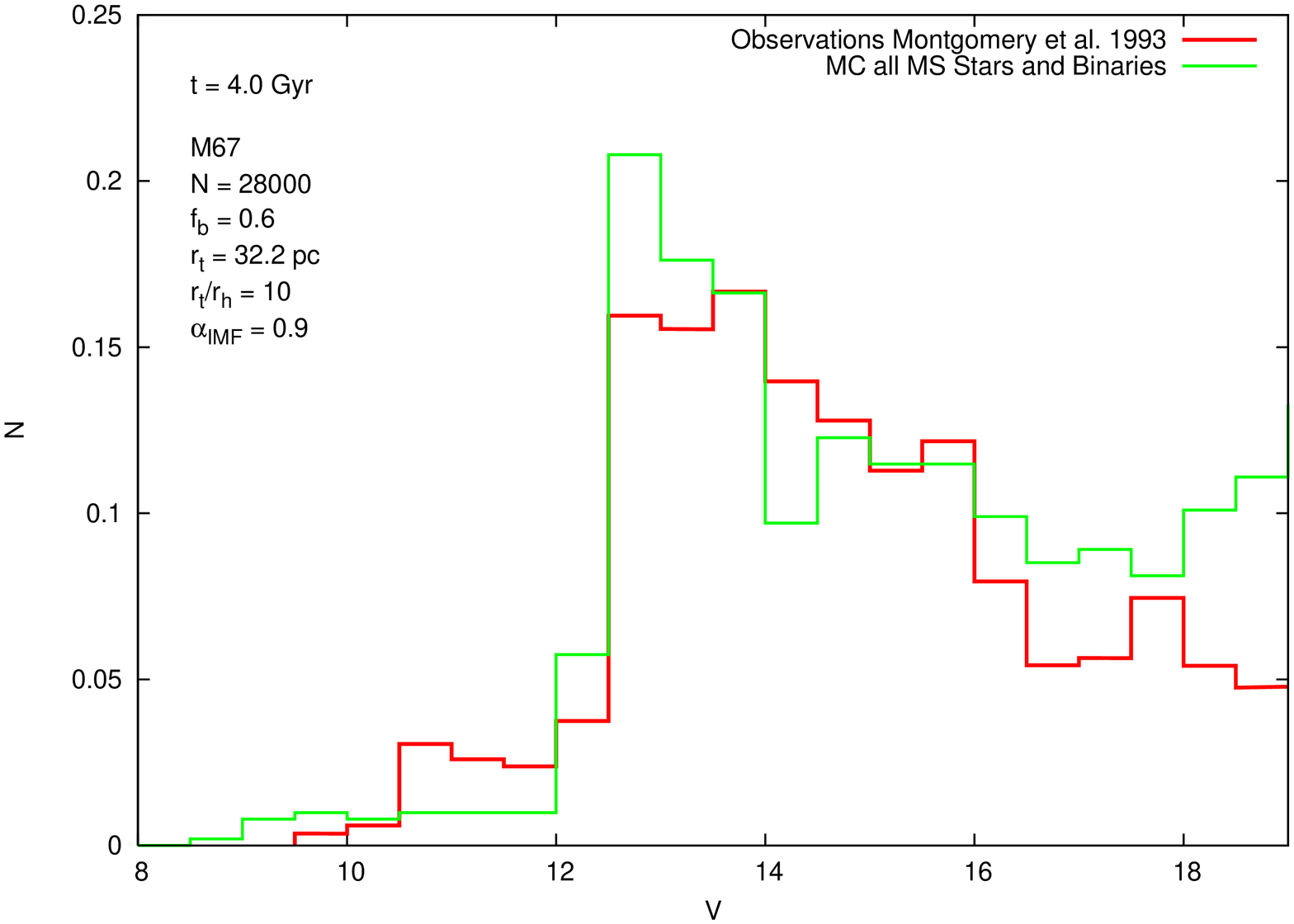}
\includegraphics[angle=-90, width=8cm]{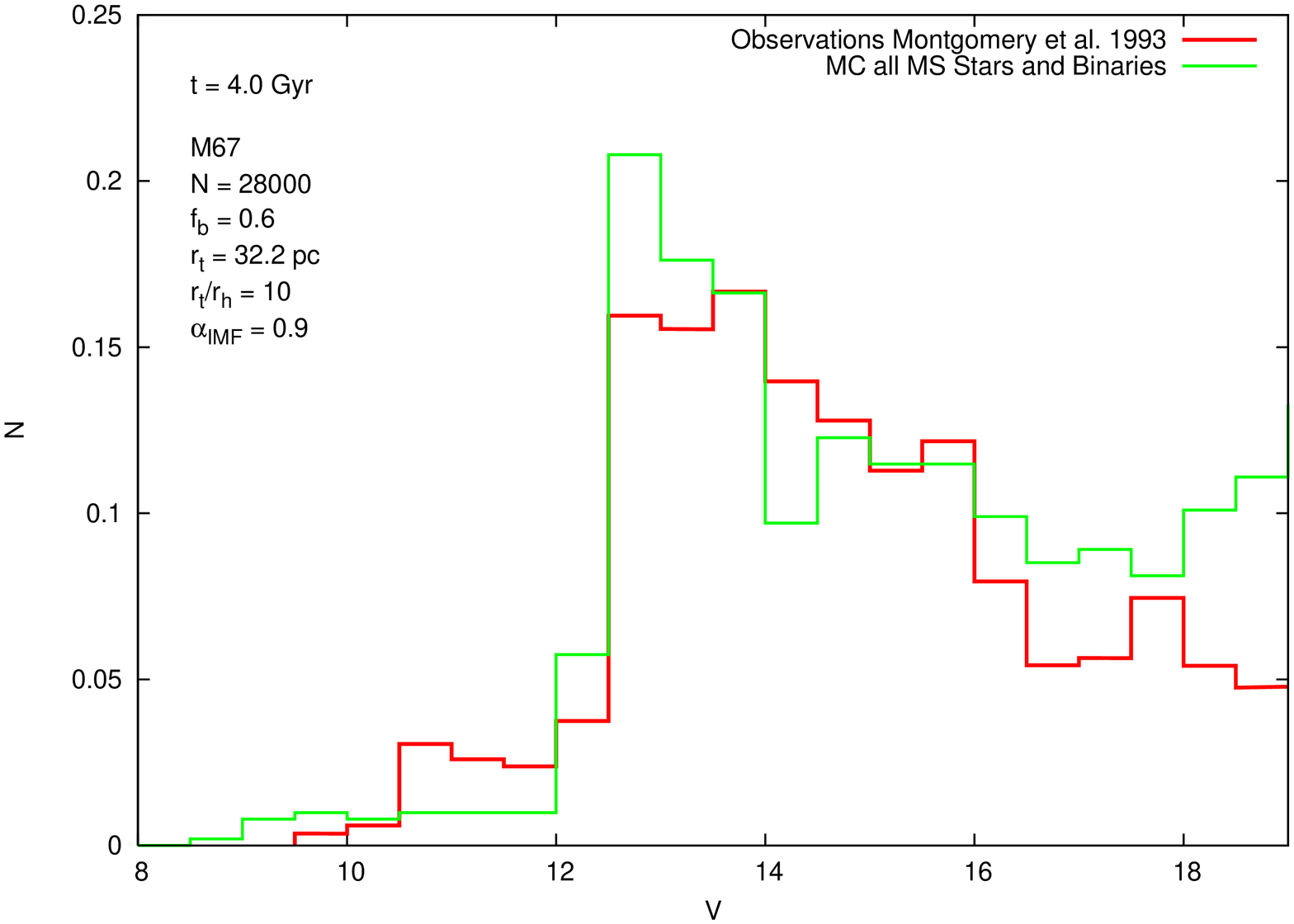}
\caption{Comparison of the surface density profile and luminosity function
for the Monte Carlo model for $\alpha_{IMF} = 0.9$ and observations. The initial 
model parameters are described in the figures. The observational data is 
described in the text}
\label{fig:m67_2000.9}
\end{figure}

\begin{figure}
\includegraphics[angle=-90, width=8cm]{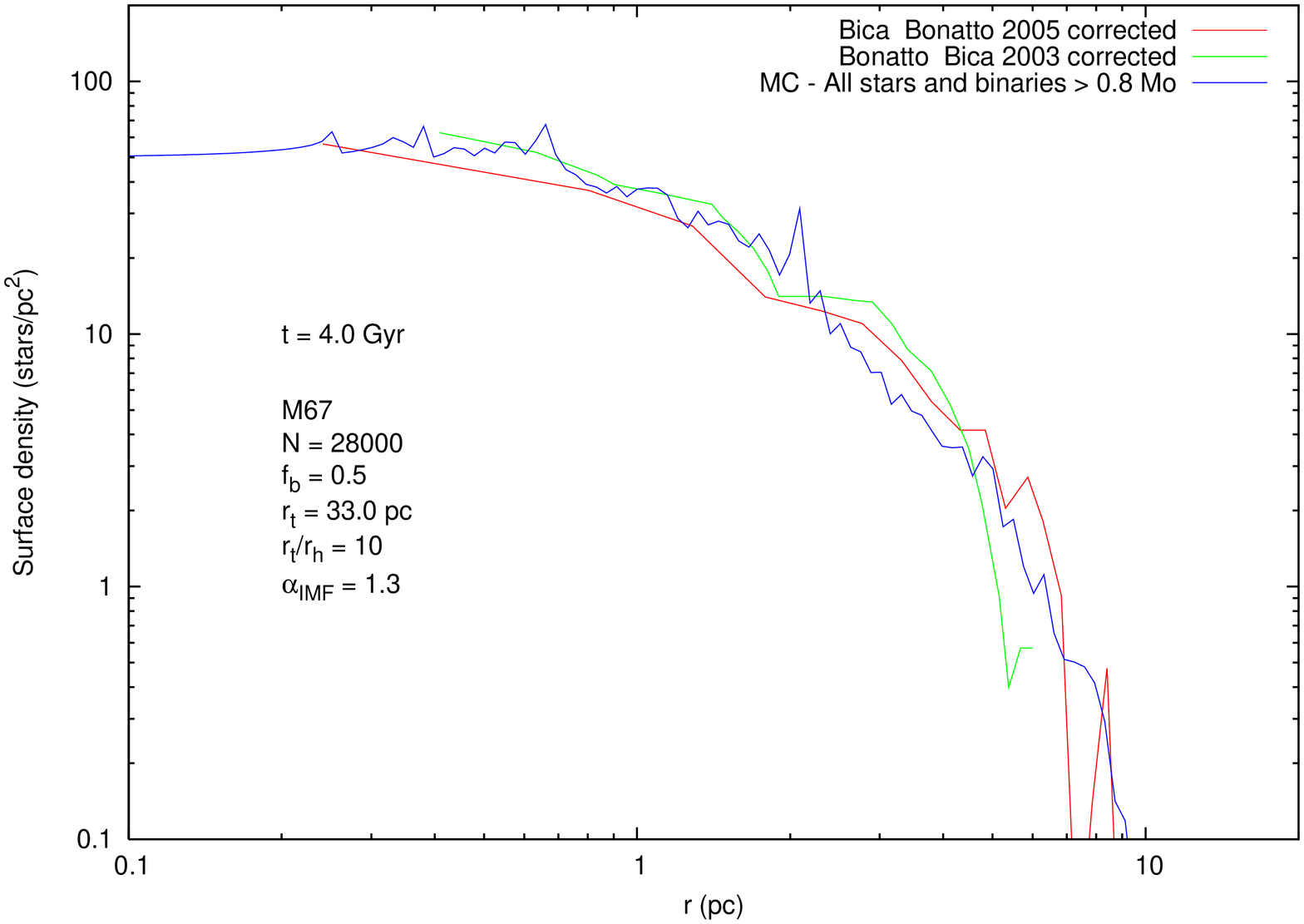}
\includegraphics[angle=-90, width=8cm]{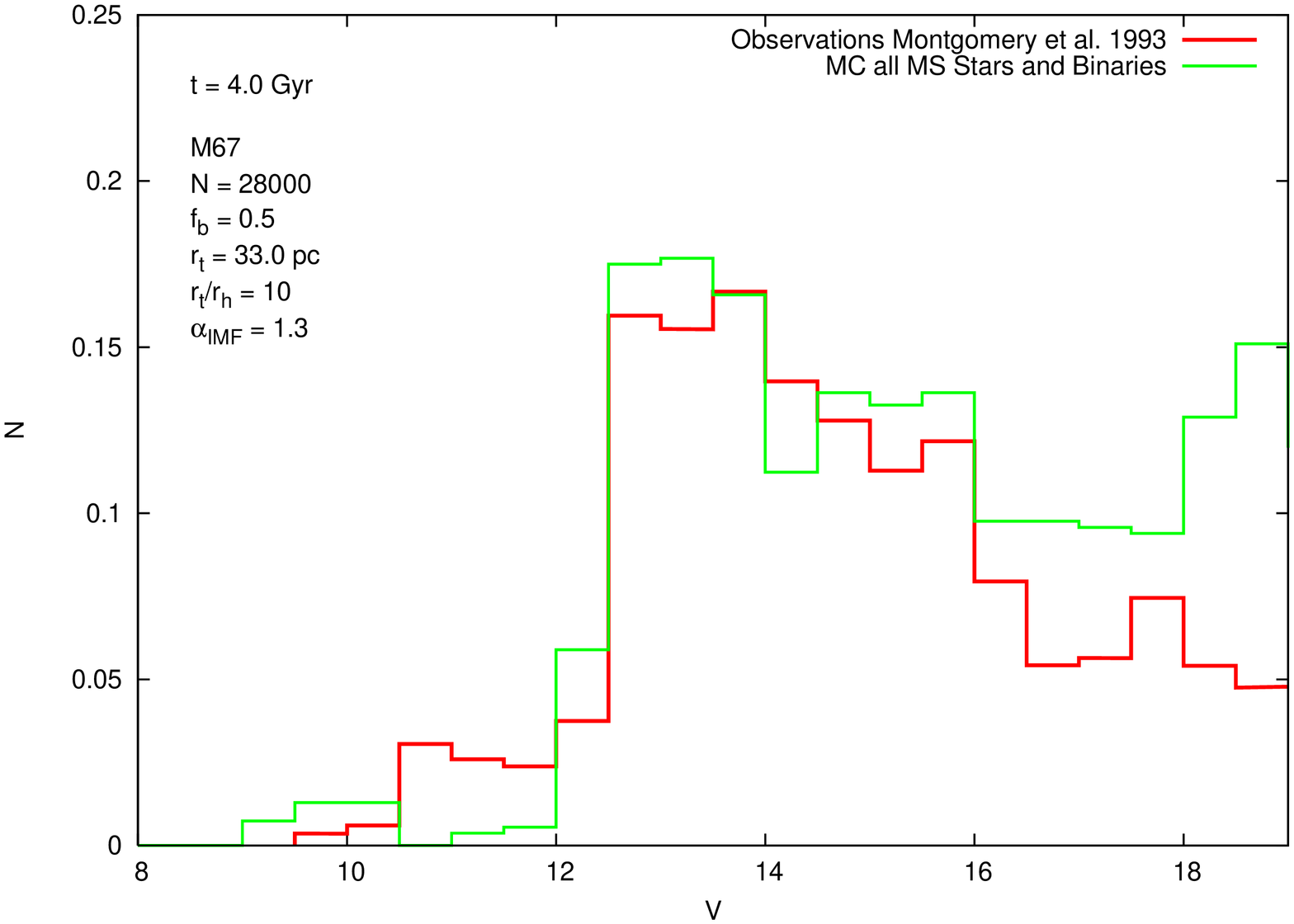}
\caption{Comparison of the surface density profile and luminosity function
for the Monte Carlo model for $\alpha_{IMF} = 1.3$ and observations. The initial 
model parameters are described in the figures. The observational data is 
described in the text}
\label{fig:m67_2001.3}
\end{figure}

Unfortunately, there is no single model which can reproduce the observational
properties of the cluster. Instead, there are several models which can 
equally well produce a reasonable fit to the observations (Figs. \ref{fig:m67_2000.1}, 
\ref{fig:m67_2000.5}, \ref{fig:m67_2000.7}, \ref{fig:m67_2000.9}, 
\ref{fig:m67_2001.3}) 
In general from very low values of $\alpha_{IMF}$, i.e. 0.1, up to the canonical 
value, 1.3, the agreement with observations is reasonably good.  For some models
the luminosity function is modelled better, while for others the surface
density profile is better. The initial parameters
of the best models and the parameters at 4 Gyr are summarised in 
Tab. \ref{table:new_m67}. The other model parameters are close to
those chosen by \citet{hurleyetal2005}. The free parameters of the Monte Carlo 
code are exactly as determined in the previous Sections.

\begin{table*}
\caption{Initial parameters, and data at time 4 Gyr, for some models of M67}
\begin{tabular}{l|llllll|llllll|l}
\hline
   & t = 0 &&&&&& t = 4 Gyr &&&&&& Figure \\
\hline
 Model & N & $M(M_{\odot})$ & $r_t$(pc) & $r_t/r_h$ & $\alpha_{IMF}$ & $f_b$ & $M(M_{\odot})$ & 
$M_{L10}(M_{\odot})$ & $r_t$ & $r_h$(pc) & $r_{hl10}$(pc) & $f_b$   \\
\hline
I   & 28000     & 28747.2 & 34.0 & 10 & 0.1 & 0.5 & 1730.6 & 1007.6 & 13.3 & 3.3 & 3.1 & 0.54 &\ref{fig:m67_2000.1}\\
II  & 28000     & 26329.2 & 32.2 & 10 & 0.5 & 0.5 & 2065.4 & 1215.0 & 13.8 & 3.2 & 3.1 & 0.53 &\ref{fig:m67_2000.5}\\
III & 32000     & 29228.5 & 32.0 & 10 & 0.7 & 0.4 & 1488.8 &  908.4 & 11.9 & 3.1 & 3.0 & 0.50 &\ref{fig:m67_2000.7}\\
IV  & 32000$^a$ & 28167.3 & 32.0 & 10 & 0.7 & 0.4 & 2815.2 & 1588.9 & 14.8 & 3.6 & 3.3 & 0.46 &\ref{fig:m67_2000.7-20}\\
V   & 28000     & 26806.7 & 32.2 & 10 & 0.9 & 0.6 & 1587.2 &  992.8 & 12.5 & 2.6 & 2.5 & 0.64 &\ref{fig:m67_2000.9}\\
VI  & 28000     & 24195.6 & 33.0 & 10 & 1.3 & 0.5 & 1799.1 & 1078.7 & 13.9 & 3.0 & 2.8 & 0.60 &\ref{fig:m67_2001.3}\\
\end{tabular}\label{table:new_m67}
\medskip

L10 -- stars with mass above $0.5 M_{\odot}$, burning nuclear fuel and contained within 10 pc

$^a$ model with different initial realisation of the sequence of random numbers. All other model
parameters are the same as for the model above.
\end{table*}

For the purpose of the following comparison with the Monte Carlo
simulations, we shall focus on the surface density profile given 
in \citet{bb2005}, corrected as above for the background stars. Clearly, 
the models show reasonably good
agreement with the observations (see Figs \ref{fig:m67_2000.1} - 
\ref{fig:m67_2001.3} (top panel)). Use of the corrected observational data brings
the surface density profile for the large radii into much better agreement with 
the simulations (see for comparison Fig. \ref{fig:m67_sd}). The special treatment 
of the tide in the Monte Carlo model plays only a minor role: the effective tidal
radius, $r_{t_{eff}}$, is only reduced by about $15 \%$ in comparison to the true
tidal radius, $r_t$. It seems that, despite the large latitude of M67, the 
contamination of the observed surface density profile by background stars plays 
an important role, at large radii.

The comparison between the luminosity functions from the Monte Carlo models and 
observations is given in the same figures as for the surface density
profiles. 
The luminosity functions are in reasonable agreement with observations, except that 
the modelled luminosity function has an excess for $V > 16$ mag, in 
comparison with observations. Of course there are also noticeable differences 
for the high-luminosity end connected with the relative lack of blue stragglers in the
Monte Carlo models. It could be argued that the excess of 
low-luminosity stars cannot be explained by supposing that not all low-mass 
stars are observed, because the observational field of M67 is high-latitude, not
heavily contaminated and sparse, and so all stars with $V \sim 18$ mag should
be observed. On the other hand the authors themselves
\citep{montgomeryetal1993} declared that the determination of the luminosity 
function ``is a difficult procedure because of the presence of background stars''.
Their background correction has been applied in the observational data shown 
in these figures, and they did it in the following way.  The background
was estimated by counting stars in an area of the colour-magnitude
diagram just blue of the main sequence, and equal in area to the
assumed boundaries of the main sequence.  But photometric errors in
the colours rise abruptly by $V = 15$, and the resulting spread in the
main sequence is very evident below $V = 18$.  If the result is that
main sequence stars spread into the area in which field stars are
counted, then the derived (observed) luminosity function will be too
small.

Because of the difficulty in quantifying this effect, we should
consider whether there are dynamical processes, omitted from the
models, which could also account for the mismatch at the faint end
of the luminosity function. The difference
in the luminosity functions for $V > 16$ mag suggests some mechanism
by which the open cluster
M67 very efficiently removed a substantial fraction of low
mass stars ($M < 0.7 M_{\odot}$) during its evolution. There are two
possibilities: 
\begin{enumerate}
\item[(i)] - removal of
residual gas during the first  few million years in a cluster
with primordial mass segregation. The resulting expansion would lead to 
preferential removal of low mass stars.
\item[(ii)] - interactions with the galactic disk 
and bulge can produce tidal shocks which in turn again preferentially removes 
low mass stars \citep{spitzer1987}. These effects would, of
course, alter the entire history of mass-loss in the models, and
require more massive initial models.
\end{enumerate}

\begin{figure}
\includegraphics[angle=-90, width=8cm]{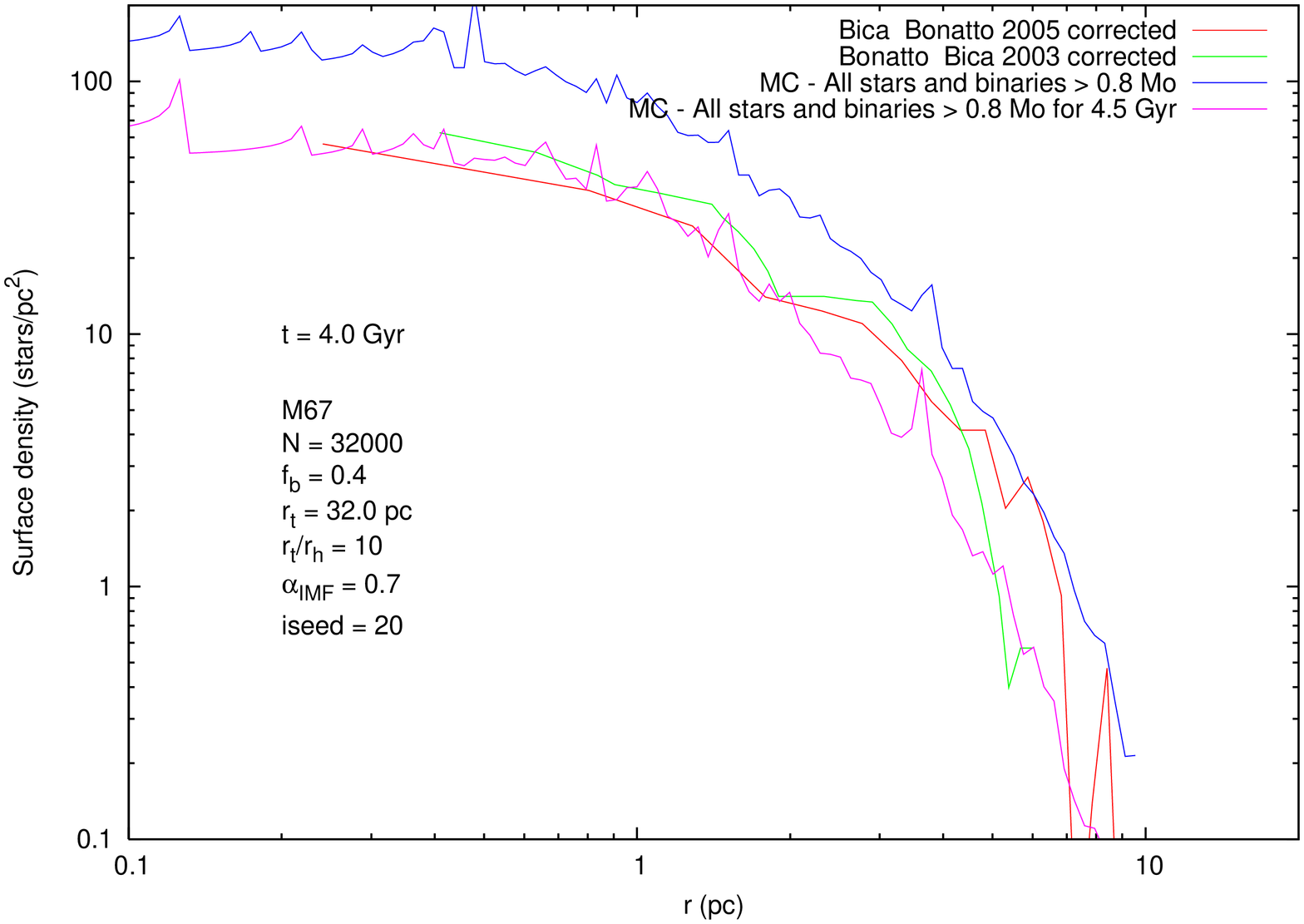}
\includegraphics[angle=-90, width=8cm]{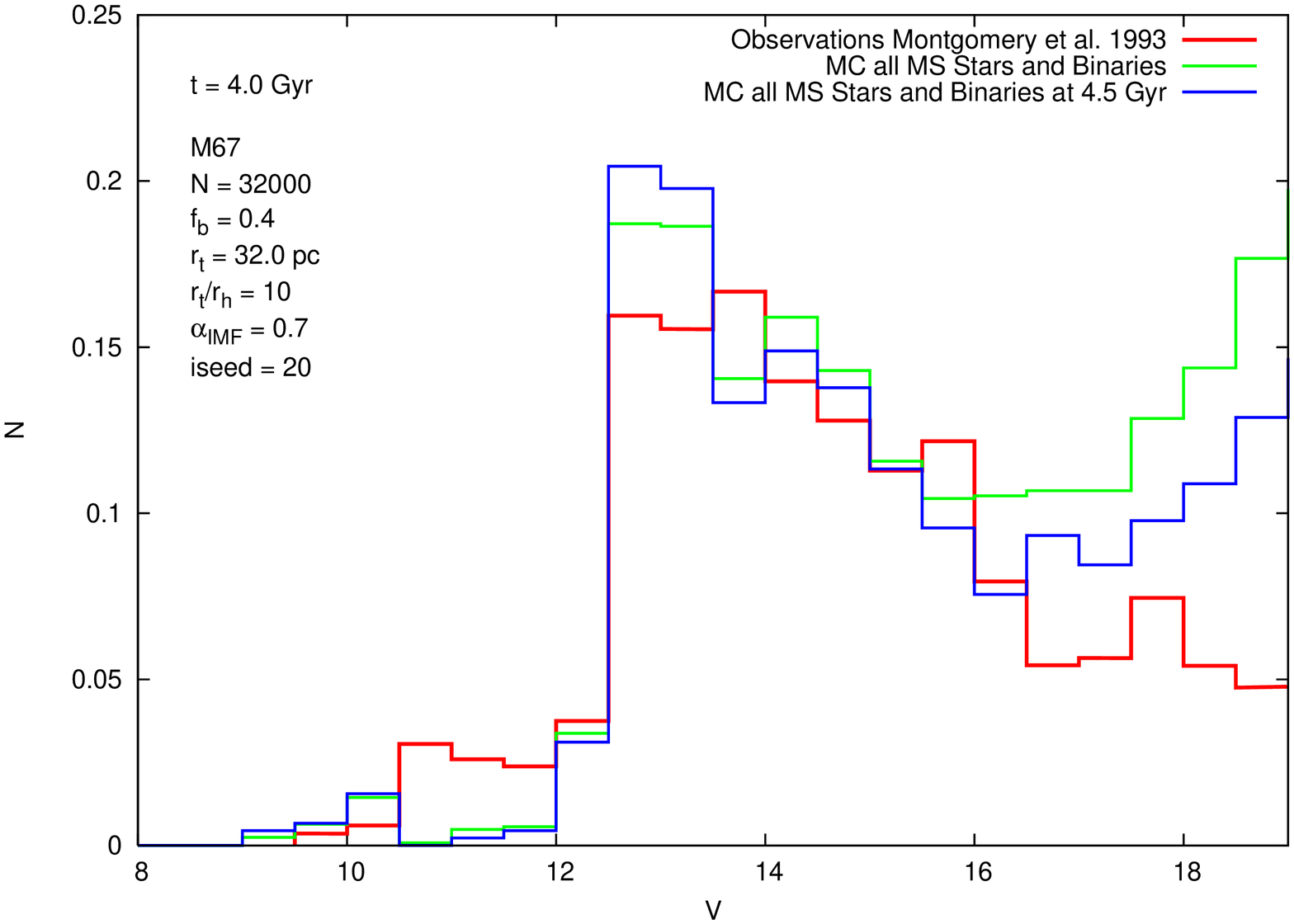}
\caption{Comparison of the surface density profile and luminosity function
for the Monte Carlo model for $\alpha_{IMF} = 0.7$ and observations for different
initial realisation of the models, iseed = 20. The initial 
model parameters are described in the figures. The observational data is 
described in the text}
\label{fig:m67_2000.7-20}
\end{figure}

Finally, we checked the influence of 
statistical fluctuations, which are intrinsic to the Monte
Carlo method, on the determination of the initial cluster parameters
which we have inferred from the comparison with observations.  To assess the
scale of this effect, the same model was repeated with different
statistical realisations of the initial model, (i.e.
different initial seeds (iseed) for the sequence of
random numbers). Typical results are presented
in Figs. \ref{fig:m67_2000.7} (Model III) and \ref{fig:m67_2000.7-20} 
(Model IV). It is clear that the use of
different realisations of the initial model has a large impact on the 
observational properties of the cluster at time 4 Gyr. The good
agreement for the surface density profile
within the half-mass radius of the cluster  is
totally destroyed: the new model (iseed = 20) has a much higher central surface 
density, by factor of about 3. Also the luminosity function for model IV shows 
poorer agreement with the observations than model III: there are many more
low-mass main sequence stars than in  model III. As can be seen 
in the Tab. \ref{table:new_m67} the model with iseed = 20 is less advanced 
in its dynamical evolution. It has larger total mass and half-mass radius, 
and a smaller binary 
fraction. As can be seen in Fig. \ref{fig:m67_2000.7-20}, at time 4.5 Gyr 
the model gives similar results to model III. Just because of the different
statistical realisation of the initial conditions, model IV has
smaller average mass than 
model III, and consequently smaller mass loss due to stellar 
evolution and a correspondingly smaller cluster expansion. The difference between 
the average masses is only about 4\%, but it seems that this is enough to change 
 the rate of cluster evolution substantially. This conclusion is supported by 
Hurley's findings \citep{hurley2007} that the incidental formation of a massive 
binary can totally change the observational properties of the cluster.
 
\section{Discussion and Conclusions}

In this paper we have presented an advanced Monte Carlo code for the evolution 
of rich star clusters, including most aspects of dynamical interactions involving 
binary and single stars, and the internal evolution of single and binary stars.  
It was shown that the free parameters of the Monte Carlo code can be 
successfully calibrated against results of small $N$-body simulations and 
 simulations of the old open cluster M67.

Sec.\ref{sec:refinement} described our best models of M67.
The results  show that an equally good fit to the observational 
data can be achieved by models which differ substantially in some initial
parameters, such as the slope of the IMF. By contrast it seems that
the other parameters are better constrained, at least within the
ensemble of models which we studied. Actually, none of the
models can successfully fit all the observational properties of M67
that we have studied, but we have argued that the remaining
mismatches can be understood in terms of known characteristics of
the Monte Carlo method, or the observational problem of subtracting
the background.  These difficulties are especially pronounced 
at the bright and faint ends of the luminosity
function. The most satisfactory models
are characterised by the following initial parameters: $N$ about
30000, $r_t$ about 33 pc, $f_b$ about 50\% and $r_t/r_h$ about 10. The word 
``about'' is used deliberately, because of the large effect of statistical 
fluctuations in such small systems.  It is worth noting that
a satisfactory fit can be achieved for a large range of
values of the power-law index of the IMF 
for low mass stars, though it seems that values of $\alpha_{IMF}$ in
the range 0.5 -- 0.7 give slightly 
better agreement with observations.  

Finally, these Monte Carlo simulations clearly show the strong influence of 
statistical fluctuations on the observational properties of a cluster.  
In consequence it seems
that recovery of the initial cluster parameters from a comparison between 
numerical models and observations may be very difficult or even impossible,
particularly for models with low or moderate $N$. More
observational data is needed to constrain the models better, but the
data on such properties as the number of particular kinds
of binaries, pulsars or blue stragglers seem
to provide only very weak constraints.         

Despite these successes in fitting $N$-body models and the
open cluster M67,
the code has some known shortcomings, which we summarise here.

\begin{enumerate}
\item {\sl Cross sections}:  it has become apparent \citep{fr2007} 
that the use of cross sections can lead to some systematic errors in the 
evolution of core parameters and other quantities.  Within the context 
of the Monte Carlo code we are working on, it is known how to replace 
these by explicit numerical calculation of the interactions 
\citep{gs2003}, and this will be our next improvement. One side effect 
of the current absence of explicit interactions is that we do not 
yet model collisions which occur during them;  therefore one channel for the 
formation of blue stragglers is missing from these simulations.
\item{\sl Higher-order multiples}:  It is widely argued that primordial 
triples and higher multiples should be incorporated into simulations along 
with primordial binaries. In any case, hierarchical triples form abundantly 
in binary-binary interactions \citep{mikkola1984}. Such higher-order 
multiples are ignored in the present Monte-Carlo code, as cross sections 
for interactions with other objects have not yet been devised.  
Hierarchical triples and higher-order multiples can be introduced as new species 
when explicit calculation of interactions has been incorporated.
\item{\sl Escape}: the Monte Carlo code described here incorporates a 
tidal cutoff, and a simple modification based on the
theory devised by \citet{baum2001}. Other treatments are possible and worth
trying.
\item{\sl Rotation}: the Monte Carlo code is based on spherical symmetry, and 
would require rather fundamental and very difficult reconstruction in order 
to cope with cluster rotation. As was pointed out by \citet{kimetal2004} the 
rotation only somewhat accelerates the rate of core collapse. 
\item{\sl Static tide}: the effect of tidal shocks have been extensively 
studied (e.g. \citet{ko1995}) and it would be possible to add the effects as 
another process altering the energies and angular momenta of the stars in the
simulations. The addition of tidal shocks will be more important when
modelling Galactic globular clusters than open clusters, which usually are 
confined inside the Galactic disk. 
\end{enumerate}

Despite these limitations, some of which are difficult to cure, the Monte Carlo
model presented in this paper shows its potential power in simulations of star
clusters, from open clusters to rich globular clusters. Monte Carlo models are
feasible in a reasonable time (a week or so) for globular clusters, which are too
large for direct $N$-body models, and future papers in this series
will present results on  M4 and several other globular clusters. 
The data provided by Monte Carlo simulations
are as detailed as those  provided by an $N$-body code. No available 
simulation methods, except Monte Carlo and $N$-body methods,
can really provide that kind
of comprehensive information. Even when $N$-body simulations eventually become 
possible, Monte Carlo models will remain as a quicker way of exploring the 
parameter space for the large scale $N$-body simulations. 

\section*{Acknowledgements}
This research was supported in part by the Polish National
Committee for Scientific Research under grant 1 P03D 002 27.  DCH
warmly thanks MG for his hospitality during a visit to Warsaw which
greatly facilitated his contribution to this project. MG warmly thanks
DCH for his hospitality during visits to Edinburgh which gave a boost to
the project. {MG also warmly thanks Janusz Kaluzny for very stimulating 
discussions about the quality of the observational data.}

\bsp

\label{lastpage}

\end{document}